\documentclass[12pt,a4paper]{article}
\usepackage[margin=2cm]{geometry}
\usepackage{comment}
\usepackage{amsmath,amssymb,extarrows,graphicx,subfigure,setspace}
\usepackage{cite}
\usepackage{slashed}
\usepackage{color}
\usepackage{braket}
\usepackage{float}
\makeatother
\usepackage{pdfpages}
\usepackage{blindtext}
\usepackage{amsfonts}
\usepackage{tensor}
\usepackage{graphicx}
\usepackage{pict2e}
\usepackage{amssymb, amsmath,environ}
\usepackage[
colorlinks=true,
linkcolor=blue,
urlcolor=blue,
filecolor=blue,
citecolor=red,
pdfstartview=FitV,
pdftitle={},
pdfauthor={},
pdfsubject={},
pdfkeywords={},
pdfpagemode=None,
bookmarksopen=true
]{hyperref}

\usepackage{epsfig}

\usepackage{hyperref}

\newcommand{\be}{\begin{equation}}
	\newcommand{\bea}{\begin{eqnarray}}
		\newcommand{\eea}{\end{eqnarray}}
	\newcommand{\ba}{\begin{array}}
		\newcommand{\ea}{\end{array}}
	\newcommand{\ee}{\end{equation}}
\newcommand{\bes}{\begin{equation*}}
	\newcommand{\beas}{\begin{eqnarray*}}
		\newcommand{\eeas}{\end{eqnarray*}}
	\newcommand{\bas}{\begin{array*}}
		\newcommand{\eas}{\end{array*}}
	\newcommand{\ees}{\end{equation*}}

\numberwithin{equation}{section}
\begin{document}
	\onehalfspacing
	\noindent
	
	\begin{titlepage}
		\vspace{10mm}

		\vspace*{20mm}
		\begin{center}
			
			{\Large {\bf Position dependence of Nielsen complexity for the Thermofield double state }\\
			}
			\vspace*{.9cm}
			
			{F. Khorasani$^{a}$, Reza Pirmoradian$^{b,d}$ and 
				M.R Tanhayi$^{a,c,\star}$
			}
			\vspace*{.5cm}
			
			$^{a}${Department of Physics, Central Tehran Branch, Islamic Azad University (IAUCTB), P.O. Box
				14676-86831, Tehran, Iran}\vspace*{5mm}
			
			${}^b $ {School of Particles and Accelerators, 	${}^c $ School of Physics, Institute for Research in Fundamental Sciences (IPM), 
				P.O. Box 19568-36681, Tehran, Iran}\vspace*{5mm}
			
			${}^d $ { Ershad Damavand, Institute for Higher Education (EDI),
				P.O. Box 14168-34311, Tehran, Iran}\\
			
			\vspace*{5mm}
			$^\star$ Email: mtanhayi@ipm.ir
			\vspace*{0.5cm}
			%{E-mails: {\tt  $^\star$mr.tanhai56@gmail.com }}%
			%{ {\tt mtanhayi@ipm.ir}}%
			\vspace*{1cm}
			%%\maketitle
		\end{center}
		
		\begin{abstract}
			In this paper, the Nielsen geometric method is used to study the position dependence of the Nielsen complexity for the thermofield double state of a harmonic oscillator. We present the state shift under the influence of an external electric field and demonstrate its importance for the construction of the corresponding circuit. By numerical analysis, we investigate the effect of the frequency and the external field on the dynamics of complexity. Our observation reveals that the system’s complexity diminishes considerably with the rise of the frequency. Furthermore, our findings indicate that the complexity exhibits a distinct behavior under a feeble external electric field, as it grows more intricate with the escalation of the frequency. However, with higher magnitudes of the electric field, the system reverts to its prior behavior. We also remark on the influence of the reference state’s frequency on the complexity.		
			
		\end{abstract}
		\newpage
		%\tableofcontents

	\end{titlepage}

	%%%%%%%%%%%%%%%%%%%%%%%%%%%%%%%%%%%%%%
	\section{Introduction}
	%%%%%%%%%%%%%%%%%%%%%%%%%%%%%%%%%%%%%%%%%%%
	Motivated by a paradigm of emerging gravity from effective quantum computation, there is an idea of the connection between information and geometry, this idea which is one of the most remarkable achievements of fundamental physics has been explored in various aspects 
	over the past ten years \cite{Swingle:2014uza}. In this regard, quantum entanglement plays a crucial role in answering the question of how spacetime emerges from quantum entanglement. According to the proposed dual correspondence between conformal field theory and classical gravity theory (AdS/CFT), the quantum entanglement between
	degrees of freedom at the boundary of geometry becomes important for the existence of bulk spacetime \cite{Kitaev:2005dm, Casini:2006es, bombelli, cw, srednicki, holo, Faulkner:2013ana, Headrick:2019eth,Chapman:2021jbh}. In the context of AdS/CFT correspondence, holographic entanglement entropy has been used as a useful measure of quantum entanglement. However, it was argued that quantum entanglement by itself is incomplete in studying some specific features of the bulk geometry. For example, a black hole's interiors may keep growing in volume essentially forever; this might be understood by the Penrose diagram. A wormhole or Einstein-Rosen bridge can connect the right and left sides of the geometry in an eternal black hole in anti-de Sitter spacetime. It is shown that the volume of the wormhole increases in time, and interestingly, the increase holds over a longer period of time which is much larger than other known time measures in the boundary theory. Considering this observation, it is argued that entanglement entropy is incomplete to capture the dynamics of the interior of a black hole  \cite{Susskind:2014moa}.  In a series of seminal papers, Susskind and his collaborators have introduced complexity at the boundary, the quantity that can probe the growth rate of the interior of the black hole \cite{Stanford:2014jda, Susskind:2014rva}. % That means black holes grow in volume because they are steadily increasing in complexity.  — an idea that, while unproven, is fueling new thinking about the quantum nature of gravity inside black holes.

	The concept of complexity was initiated by theoretical computer science, and it indicates a measure of the difficulty of performing a proposed task. In other words, the computational complexity is the minimum number of operations needed to get a desired (target) state from an initial (reference) state. Based on the holographic principle, Susskind used quantum complexity as a probe of the growth of the wormhole inside a black hole within two proposals: complexity$=$volume \cite{Susskind:2014rva, Stanford:2014jda} and 
	complexity$=$action \cite{Brown:2015lvg}. According to the first proposal, the quantum complexity on the boundary becomes dual to the maximum volume of the Einstein-Rosen Bridge in the bulk spacetime  
	while in the second proposal, the complexity is related to the
	gravitational action computed on a specific region of the Wheeler-De Witt patch defined in the bulk spacetime.\footnote{For more details see Ref \cite{Belin:2021bga}. } It is worth mentioning that in quantum field theory, due to the difficulty of computation, many aspects of quantifying complexity have not been yet explored (Ref.\cite{Ali:2018fcz} and references therein, see also  \cite{Alishahiha:2019cib, Engelhardt:2021mju, RezaTanhayi:2018cyv, Brown:2021rmz,Alishahiha:2022nhe, Alishahiha:2015rta, Alishahiha:2018lfv, Borvayeh:2020yip}). For quantum systems with gravity dual, by defining the thermofield double (TFD) states which are dual to an eternal black hole in anti-de Sitter spacetime, it is argued that such states could provide a powerful setup for investigating various features of black hole physics, as well as, the time evolution of entanglement, complexity and, etc \cite{Cottrell:2018ash, Chapman:2018hou}.  By definition, for two given copies of any quantum mechanical system, one can define the thermofield double state as follows 
	\bea\label{TFD.11}  
	\mid\hspace{-1mm}  \text{TFD}\rangle\equiv\frac{1}{\sqrt{Z}}\displaystyle\sum_{n=0}^{\infty}e^{-\frac{\beta}{2}\mathcal{E}_n}\mid\hspace{-1mm}  \mathcal{E}_{n}\rangle\hspace{1mm}\mid\hspace{-1mm}  \mathcal{E}_{\tilde{n}}\rangle,
	\eea
	note that in writing equation \eqref{TFD.11}, the energy basis is being used. This is the unique pure state and each of the two copies has its own temperature $\beta^{-1} $ and $Z$ is the canonical partition function. %The product space of the two systems is defined by  \bea  \mid\hspace{-1mm} n,\tilde{m}\rangle =\mid\hspace{-1mm} \mathcal{E}_{n}\rangle\hspace{1mm} \mid\mathcal{E}_{\tilde{m}}\rangle,  \eea
	The tilde system is supposed to be an identical copy of the original system. Making use of the covariance matrix approach, for the TFD state, the entanglement entropy and 
	logarithmic negativity have been studied in \cite{Ghasemi:2021jiy}, moreover, Ref. \cite{Doroudiani:2019llj} deals with the complexity of a charged TFD state in the presence of an electric field. In this paper, we extend the results of Ref. \cite{Doroudiani:2019llj} in parts, and for a simple quantum system, namely a harmonic oscillator,  we construct the appropriate TFD state and examine the complexity dynamics by altering one side of the state, which is done by applying a weak external electric field. The quantum complexity calculations in this paper, are based on  Nielsen's geometric approach\cite{Nielsen:2005,Nielsen:2006,Nielsen:2007, DiGiulio:2020hlz}, and for Gaussian states, we use the covariance matrix approach to calculate the complexity of the state. 
	
	The layout of this paper is as follows. Section 2 revisits the construction of a TFD state for a harmonic oscillator. Next, we establish the target and reference states for two coupled harmonic oscillators and investigate the complexity of transforming the reference state into the target state. Section 3 recapitulates the covariance matrix method and applies it to derive the complexity and examine the complexification of the suggested states by altering the parameters of the theory, such as the external electric field and the frequency. The concluding remarks are given in section 4. In Appendix \ref{appA}, we give some details of the decomposition of TFD on a diagonal basis. In Appendix \ref{appB}, we define the reference and target states and build a circuit with gate operators that give us the target state. Finally, in Appendix \ref{appC}, we find the eigenvalues of the covariance matrix. 
	
	%%%%%%%%%%%%%%%%%%%%%%%%%%%%%%%%%%%%%%%%%%%%%%%%%%%%%%%%%%
	
	\section{Thermofield double states for a harmonic oscillator with an external electric field  }
	%%%%%%%%%%%%%%%%%%%%%%%%%%%%%%%%%%%%%%%%%%%%%%%%%%%%%%%
	An ensemble of harmonic oscillators is often used to discretize theories on a spatial lattice, e.g. Ref. \cite{Jefferson:2017sdb} used this method to study the ground state complexity of free bosonic quantum fields. Considering the importance of the TFD state, in Ref. \cite{Chapman:2018hou}, for a simple harmonic oscillator the complexity and entanglement entropy for a TFD state have been studied. Moreover,  Ref.\cite{Doroudiani:2019llj} considered the charged TFD state, and in this section, we want to obtain a TFD state for a harmonic oscillator in the presence of an external electric field. % This can be achieved in two different ways: One way is simply using the exact form of the energy eigenstates of the harmonic oscillator, and in the second approach, one can interpret the energy eigenstate of the harmonic oscillator as a state composed of $n$ particles, each with charge $q$, and then put this system of charged particles in an electric field. 
	The Hamiltonian of a harmonic oscillator in a constant external electric field $E$,  is given by 
	\bea\label{HE}
	\hat{H}=\frac{\hat p^2}{2m}+\frac{1}{2}m\omega^2(\hat x-\mathbf{d})^2-\mathcal{E},
	\eea
	where
	\bea\label{Eq}
	\mathbf{d}=\frac{q E}{m\omega^2},\hspace{1cm}\mathcal{E}=\frac{q^2 E^2}{2m\omega^2}.
	\eea
	The energy levels and the normalized energy eigenstates are as follows  
	\bea\label{En1}
	\mid\hspace{-1mm} \mathcal{E}_n\rangle=\sqrt{\frac{1}{n!}}\hspace{.5mm}(\hat{a}^\dagger)^n \mid\hspace{-1mm}  0\rangle,\,\,\,\,\,\,\,\,\,		\mathcal{E}_{n}\equiv \mathcal{E}_n^{(0)}-\mathcal{E}= \omega(n+\frac{1}{2})-\mathcal{E} 
	\eea
	where the modified canonical creation and annihilation operators are given by
	\bea\label{aq}
	\hat{a}^\dagger=\sqrt{\frac{m\omega}{2}}\left( (\hat{x}-\mathbf{d})-i\frac{\hat p}{m\omega}\right),\hspace{1cm}\hat{a}=\sqrt{\frac{m\omega}{2}}\left((\hat{x}-\mathbf{d})+i\frac{\hat p}{m\omega}\right).
	\eea
	As mentioned in the introduction, for our case we shall construct the corresponding TFD state, where at $t=0$ and by making use of \eqref{TFD.11} one can write
	\bea\label{TFD.1}  
	\mid\hspace{-1mm}  \text{TFD}\rangle&=&\frac{1}{\sqrt{Z}}\displaystyle\sum_{n=0}^{\infty}e^{-\frac{\beta}{2}\left(\mathcal{E}_n^{(0)}-\mathcal{E}\right)}\mid\hspace{-1mm}  \mathcal{E}_{n}\rangle\hspace{1mm}\mid\hspace{-1mm}  \mathcal{E}_{\tilde{n}}\rangle\nonumber\\
	&=&\frac{1}{\sqrt{Z}}\displaystyle\sum_{n=0}^{\infty}e^{-\frac{\beta}{2}\left(\mathcal{E}_n^{(0)}-\mathcal{E}\right)}\hspace{1mm}\frac{1}{ n!}\hspace{.5mm}(\hat{a}^\dagger_{L})^n(\hat{a}^\dagger_{R})^n \mid\hspace{-1mm}  0\rangle_L\mid\hspace{-1mm} \mid\hspace{-1mm} 0\rangle_R,
	\eea
	where the subscript $L$ and $R$ stand for the left and right quantum theories. To find the normalization factor, let us compute the inner product of the state as follows
	\bea
	\begin{split}
		\langle \text{TFD}\hspace{-1mm}\mid\hspace{-1mm}  \text{TFD}\hspace{-.1mm}\rangle &=\frac{1}{Z}\displaystyle\sum_{n,n^{\prime}=0}^{\infty}e^{-\frac{\beta}{2}\left(\mathcal{E}_n^{(0)}+\mathcal{E}_{n^{\prime}}^{(0)}-2\mathcal{E}\right)}\hspace{1mm} \bigg(\hspace{1mm}_{L}\langle \mathcal{E}_{n^{\prime}}\hspace{-1mm}\mid\hspace{1mm}_{R}\langle \mathcal{E}_{n^{\prime}}\hspace{-1mm}\mid\bigg)\bigg(\mid\hspace{-1mm} \mathcal{E}_{n}\rangle_{L}\hspace{1mm}\mid\hspace{-1mm} \mathcal{E}_{n}\rangle_{R}\bigg)\\&=\frac{1}{Z}\displaystyle\sum_{n=0}^{\infty}e^{-\beta\left(\mathcal{E}_n^{(0)}-\mathcal{E}\right)}=\frac{1}{Z}e^{\beta\mathcal{E}}\displaystyle\sum_{n=0}^{\infty}e^{-(n+\frac{1}{2})\beta\omega}\\ &
		= \frac{1}{Z} \frac{e^{\beta\mathcal{E}}}{e^{\frac{\beta \omega}{2}}-e^{- \frac{\beta\omega}{2}}},
	\end{split}
	\eea
	leading to the following expression  
	\bea
	Z=\frac{e^{\beta\mathcal{E}}}{e^{ \frac{\beta\omega}{2}}-e^{- \frac{\beta\omega}{2}}}\hspace{.5mm}.
	\eea
	Consequently, the thermofield double state (\ref{TFD.1}) becomes
	\bea\label{TFD.2}
	\mid\hspace{-1mm} \text{TFD}\rangle
	= \sqrt{1-e^{-\beta\omega}}\hspace{1mm}\exp{\left(e^{-\frac{\beta \omega}{2}}\hspace{1mm}\hat{a}^\dagger_{L}\hat{a}^\dagger_{R}\right)}\mid\hspace{-1mm}0\rangle_L\hspace{.5mm}\mid \hspace{-1mm}0\rangle_R,
	\eea
	where $\hat{a}^\dagger$ is defined by \eqref{aq}. The energy level shifts due to the electric field, which is equivalent to adding a constant term to the energy value (equation \eqref{En1}), and the TFD state \eqref{TFD.2} remains unaffected by this, it does not include the shift\footnote{We appreciate the referee's insightful explanations, which prompted us to revise the presentation of our article.}. 	This work aims to evaluate how the complexity varies with the coordinate, and  to explore this issue, we use a normal basis that is defined by	
	\bea\label{xpm}
	\hat{x} _\pm=\frac{1}{\sqrt{2}}\Big(\hat{x}_L\pm \hat{x}_R\Big),\hspace{.9cm}
	\hat{p}_\pm=\frac{1}{\sqrt{2}}\Big(\hat{p}_L\pm \hat{p}_R\Big),
	\eea	 
	therefore, the  state \eqref{TFD.2} decomposes as  (Appendix \ref{appA})
	\bea\label{Opm}
	\mid \hspace{-1mm}\Psi_T\rangle= e^{-i\alpha(\hat{x}_+\hat{p}_+-\sqrt{2}\mathbf{d} \hat{p}_+)}\, e^{i\alpha\hat{x}_-\hat{p}_-}\hspace{-1mm}\mid\hspace{-1mm}0\rangle_+\,\mid\hspace{-1mm}0\rangle_-,
	\eea
	where we have defined
	\bea\label{al} \alpha=\cosh^{-1}\frac{1}{\sqrt{1-e^{-\beta\omega}}}.
	\eea
	Moreover, in order to study the dynamics of a system, it is also necessary to know the time evolution of the corresponding state, 
	\bea
	\mid\hspace{-1mm} \Psi_{T}(t)\rangle=(1-e^{-\beta\omega})^{\frac{1}{2}}\displaystyle\sum_{n=0}^{\infty}e^{-n\frac{\beta\omega}{2}}e^{-i\left[(n+\frac{1}{2})\omega-\mathcal{E}\right]t_L}e^{-i\left[(n+\frac{1}{2})\omega-\mathcal{E}\right]t_R}\hspace{-1mm}\mid\hspace{-1mm} \mathcal{E}_{n}\rangle_L\mid\hspace{-1mm} \mathcal{E}_{n}\rangle_R,
	\eea
	it is supposed that the system evolves with time,  $t = t_L + t_R$ which allows us to consider the symmetric times $t_L = t_R = t/2$. This implies
	\bea\label{TTFD.1}
	\begin{split}
		\mid \hspace{-1mm}\Psi_{T}(t)\rangle&=e^{-i(\frac{\omega}{2}-\mathcal{E})t}(1-e^{-\beta\omega})^{\frac{1}{2}}\displaystyle\sum_{n=0}^{\infty}e^{-n\frac{\beta\omega}{2}}e^{-in\omega t}\mid\hspace{-1mm} \mathcal{E}_{n}\rangle_L\mid\hspace{-1mm} \mathcal{E}_{n}\rangle_R \\&=e^{-i(\frac{\omega}{2}-\mathcal{E})t}(1-e^{-\beta\omega})^{\frac{1}{2}}\displaystyle\sum_{n=0}^{\infty}\frac{e^{-n\frac{\beta\omega}{2}}e^{-in\omega t}}{n!}(\hat{a}^\dagger_{L}\hspace{.5mm}\hat{a}^\dagger_{R})^{n}\mid\hspace{-1mm}0\rangle_L\mid\hspace{-1mm}0\rangle_R\\&=e^{-i(\frac{\omega}{2}-\mathcal{E})t}(1-e^{-\beta\omega})^{\frac{1}{2}}\hspace{1mm}\exp\bigg[{e^{-\frac{\beta\omega}{2}}e^{-i\omega t}(\hat{a}^\dagger_{L}\hspace{.5mm}\hat{a}^\dagger_{R})}\bigg]\hspace{1mm}\mid\hspace{-1mm}0\rangle_L\mid\hspace{-1mm}0\rangle_R.
	\end{split}
	\eea
	Making use of the same steps as done in appendix \ref{appA}, the time-dependent state can be written as 
	\bea\label{TTFD.2} \mid\hspace{-1mm} \Psi_{T}(t)\rangle=e^{-i(\frac{\omega}{2}-\mathcal{E})t}\exp\bigg[\alpha e^{-i\omega t}\hspace{.5mm}\hat{a}^\dagger_{L}\hspace{.5mm}\hat{a}^\dagger_{R}-\alpha e^{i \omega t}\hspace{.5mm}\hat{a}_{L}\hspace{.5mm}\hat{a}_{R}\bigg]\hspace{-1mm}\mid\hspace{-1mm}0\rangle_L\,\mid\hspace{-1mm}0\rangle_R. \eea 
	By dropping the global phase (it does not play any role in studying the complexity of the state), the above state on the diagonal basis can be simplified as 
	\bea\label{TCTFD.3}
	\mid\hspace{-1mm} \Psi_{T}(t)\rangle= \exp\Big[-i\alpha\hat{\mathcal{D}}_{+}(t)+i\alpha\hat{\mathcal{D}}_{-}(t)\Big] \hspace{-.5mm}\mid\hspace{-1mm}0\rangle_+  \hspace{-.5mm}\mid\hspace{-1mm}0\rangle_-
	\eea
	where we have defined 
	\bea
	&&\hat{\mathcal{D}}_{-}(t)= \frac{1}{2}\Big[\cos\omega t\hspace{.5mm}(\hat{x}_- \hat{p}_- + \hat{p}_- \hat{x}_-)+\sin\omega t\hspace{.5mm}(m\omega{\hat{x}}^2_--\frac{1}{m\omega}\hat{p}^2_-)\Big],
	\cr\nonumber\\
	&&\hat{\mathcal{D}}_{+}(t)=\frac{1}{2}\Big[\cos\omega t\left(\hat{x}_+ \hat{p}_+ +\hat{p}_+ \hat{x}_+ -2\sqrt{2}\hspace{.5mm}\mathbf{d} \,\hat{p}_+ \right)
	\cr\nonumber\\
	&&\hspace{3.5cm} +\sin\omega t\left(m\omega\,\hat{x}_+^2 -\frac{1}{m\omega}\hat{p}^2_+-2\sqrt{2}m\omega\hspace{.5mm}\mathbf{d}\,\hat{x}_++2m\omega\,\mathbf{d}^2\right)\Big]\nonumber.
	\eea
	We would like to mention that the above time-dependent state is the extension of the previous works (Ref.\cite{Doroudiani:2019llj} and equation (36) in Ref. \cite{Chapman:2018hou}), in which, in deriving \eqref{TCTFD.3}, we have investigated the effect of an external electric field that can be viewed as a displacement in the position due to the fixed gates employed to generate the shifted TFD state.  Subsequently, we explore the position dependency of the complexity arising from this displacement. In the following section, we shall study the complexification of the above state, and to do so, we first recall the covariance matrix approach. 
	
	%%%%%%%%%%%%%%%%%%%%%%%%%%%%%%%%%%%%%%%%%%%%%%%%%%%
	\section{Circuit Complexity: the covariance matrix approach}\label{comatrix}
	%%%%%%%%%%%%%%%%%%%%%%%%%%%%%%%%%%%%%%%%%%%%%%%%%%
	%In this section, by making use of the covariance matrix approach, we would like to find the complexity of time-dependent state (\ref{TCTFD.3}). 
	In the literature, there are different but equivalent definitions of complexity, e.g., it is defined as the minimum number of gates to execute a unitary transformation where it must transform the reference state to the desired output target state. A unitary operator $\hat{ U}$ must be found to build the target state  $\mid\hspace{-1mm}\Psi_{T}\rangle$ from the reference state $\mid\hspace{-1mm}\Psi_{R}\rangle$ as %There is albeit an error of $\epsilon$ in this construction. We note that the reference state is not an entangled state as long as it is a factorizable state in position space.  In Nielsen's geometric approach  \cite{Nielsen:2005,Nielsen:2006,Nielsen:2007}, the optimal unitary transformation $\hat{U}$ between reference and target state is defined as follows
	\bea\label{u}
	\mid\hspace{-1mm}\Psi_{T}\rangle=\hat{ U} \mid\hspace{-1mm}\Psi_{R}\rangle.
	\eea
	In this section for Gaussian states, we use a covariance matrix approach, first briefly review this approach, and then use this method to consider the complexity of the TFD state for a harmonic oscillator.\\ In fact,  $\hat U$ is a sequence of continuous elementary unit gates, called a circuit,  Of course, one can define so many circuits from the reference state to the target state, and they can be parameterized by 
	\bea\label{U}
	{\cal U}(s) = \mathcal{P}\hspace{1mm} \exp\Big[-i \int_{0}^{s}ds'\,Y(s')\,\hat{\mathcal{O}} \hspace{1mm} \Big ],
	\eea
	noting that in the space of these circuits (paths), the following boundary condition 
	$${\cal U}(s=0)=\mathbb{I}\,\,\,\,\,\,\,\mbox{and}\,\,\,\,\,\,\,{\cal U}(s=1)=\hat U$$
	gives us the desired $\hat U$ operator. In \eqref{U}, $\mathcal{P}$ stands for the path ordering operator, $s$ is an affine parameter that labels the paths; $\hat{ {\mathcal{O}}}$ is a set of generators dual to each gate, and $Y(s)$ is the control function that identifies the number of gates at each path. The problem is to find the shortest path (optimal circuit) that leads to the target state among an infinite number of paths, which can be achieved through trajectory optimization. In this sense, the cost function $F$ is defined for each path as a function of ${\cal U}$ and its corresponding control function that parameterizes the costs of the various pathways as follows \cite{Chapman:2018hou}
	\bea\label{costf}
	\mathcal{C}({\cal U}) = \int _{0}^{1} ds \hspace{1mm} F\Big({\cal U}(s),Y(s)\Big)
	\eea 
	Therefore, complexity is, by definition, the lowest effort required to produce a target state from a given reference state; The functional \eqref{costf} must be minimized, which means that any trajectory that can lead from the reference state to the construction of the target state must be optimized. In other words, the complexity of the desired unitary operator is 
	\begin{equation}
		\cal C=\min {\cal C}(\cal U),\hspace{.4cm}\mbox{over all possible Y }	
	\end{equation} 
	Let the cost function satisfy some conditions, such as continuity, positivity, homogeneity, and triangle inequality. These four properties are nearly sufficient to define a class of geometries known as Finsler manifolds. Then, the equation ${\cal U}(s)$ defines the length functional of any curve with a Finsler metric, and $ Y(s)$  becomes the tangent vector to the trajectory (for more details, see \cite{Chapman:2018hou})
	\be Y(s)\,\,\hat{\mathcal{O}}=\frac{d\, {\cal U}(s)}{d t}\,\, {\cal U}^{-1}(s).\ee 
	Neilsen has formulated the problem of finding an optimal circuit as the problem of finding extremal curves, or geodesics, in the Finsler geometry, and the complexity is then associated with the length of the geodesic. In principle, in this notation, there are some cost functions, and, in this paper, we use the following relation for the cost function 
	$$F({\cal U}(s),Y(s))=\sqrt{\sum_{I} \mid\hspace{-1mm} Y^{I}\hspace{-1mm}\mid^2}\,\,.$$  
	The minimization is a difficult task, however, for the Gaussian states, the standard method of studying trajectories is the covariance matrix formalism. The covariance matrix for a pure Gaussian state is the two-point function defined by 
	\bea
	&& G = \langle\Psi\hspace{-1mm}\mid\hat{\xi}^{a}\hat{\xi}^{b}+\hat{\xi}^{b}\hat{\xi}^a \mid\hspace{-1mm}\Psi\rangle
	\eea
	where $\hat{\xi} = \{\hat{x},\hat{p}, {\mathbf{d}} \}$ in which we have defined ${\mathbf{d}}\,(=\mathbf{d} \,\mathbb{I}$) and we demonstrate its role in gate optimization in a specific circuit\footnote{In Appendix \ref{appB}, we investigate this point by constructing a circuit.}. Each Gaussian state can be fully characterized by its covariance matrix. In addition, trajectories between states can also be easily identified in the state space using the covariance matrix \cite{Chapman:2018hou}. 
	In the space of Gaussian states one can identify the unitary operator $\hat{U}(s)$ that maps two states as 
	\bea
	\mid\hspace{-1mm}\Psi_{\text{s}}\rangle = \hat{U}(s)\mid\hspace{-1mm}\Psi_{\text{0}}\rangle,
	\eea
	where $\hat{U}(s)$ can be written in terms of  a Hermitian operator as follows 
	\bea\label{uu}
	\hat{U}(s) \equiv e^{-is\hat {K}}= \exp[{- s\frac{i}{2}\hat{\xi}^{a}\hspace{.5mm} k_{ab}\hspace{.5mm}\hat{\xi}^{b}}],
	\eea
	where  $\hat{K}$ is the Hermitian operator and its matrix elements are denoted by $k_{ab}$. The covariance matrix for $\mid\hspace{-1mm}\Psi_{\text{s}}\rangle$ can be computed as follows
	\bea\label{GsG0}
	&& G_{s} = \langle\Psi_{\text{s}}\hspace{-1mm}\mid\hat{\xi}^{a}\hat{\xi}^{b}+\hat{\xi}^{b}\hat{\xi}^a\mid\hspace{-1mm}\Psi_{s}\rangle
	={U}(s) \hspace{1mm}G_{0} \hspace{1mm} {U}(s)^\intercal,
	\eea
	noting that $U(s)$ is the matrix representation of the unitary operator $\hat{U}(s)$   \bea \hat{U}^{\dagger}(s)\hspace{.5mm}\hat{\xi}^{a}\hspace{.5mm}\hat{U}(s) =  U(s)^{a}_{b}\hspace{1mm}\hat{\xi}^{b}. \eea
	Given that the reference and target states have covariance matrices $G_R$ and $G_T$, respectively, where  \begin{equation}
		G_T={U}(s=1) \hspace{1mm}G_{0} \hspace{1mm} {U}(s=1)^\intercal,
	\end{equation} 
it is argued that the information on complexity can be deduced from the relative covariance matrix which is defined as  \cite{Chapman:2018hou}
	\bea\label{Delta}
	{\cal M} = G_{T} \hspace{1mm}(G_{R})^{-1},
	\eea  
	and the complexity, in a way that depends on the eigenvalues of the above-mentioned matrix, is given by 
	\bea\label{ck2}
	\mathcal{C}  =\frac{1}{4}\text{Tr} \left[\mid\hspace{-.5mm}\log	{\cal M}\hspace{-.6mm}\mid^2\right].
	\eea 
 We should mention that in (\ref{uu}) the $\hat{K}$ operators consist of all quadratic combinations of the dimension-full operators $\hat{\xi}$. This means that the control functions $Y^I$ are dimension-full and to sum them up and obtain appropriate cost functions, they must be dimensionless. The simplest approach is introducing new dimensionless position and momentum coordinates by introducing a new gate scale $g_{s}$, as follows
	\bea\label{xpnew}
	\hat{\xi}_{\text{new}} = \{g_s \hat{x},\,\frac{\hat{p}}{g_s},\, g_s\hspace{.2mm} {\mathbf{d}}\}.
	\eea
	Introducing the $g_s$ factor makes the new coordinates dimensionless. To calculate the complexity, we rewrite the state (\ref{TCTFD.3}) as follows
	\bea\label{TCTFD.4}
	 \Psi(t)= {U}_+(s=1)\,(G_0)_+\otimes {U}_-(s=1) (G_{0})_-=e^{-i{K}_+}(G_0)_+ \otimes e^{-i{K}_-}(G_0)_-,
	\eea
	 represents two different modes associated with the harmonic oscillators that move to the right and left which are decomposed.
	
	Now, we use the method reviewed above to study the complexity of the state (\ref{TCTFD.3}). 
	With the new coordinate $\hat{\xi}_{\text{new}} $ and making use of    \eqref{TCTFD.3}, \eqref{uu}, the matrices $(k_+)_{ab}$ and $(k_-)_{ab}$ are found to be
	\bea 
	(k_-)_{ab}=
	-\alpha\begin{pmatrix}
		\lambda\sin\omega t & \cos\omega t  & 0 \\
		\cos\omega t &  -\frac{1}{ \lambda}\hspace{.5mm}\sin\omega t & 0  \\
		0 & 0  & 0 
	\end{pmatrix}
	\eea
	and
	\bea 
	(k_+)_{ab}=
	-\alpha\begin{pmatrix}
		-\lambda\sin\omega t & -\cos\omega t  &\sqrt{2}\hspace{.5mm}\hspace{.5mm}\tilde{\mathbf{d}}\hspace{.5mm}\lambda \sin\omega t \\
		-\cos\omega t &  \frac{1}{ \lambda}\sin\omega t & \sqrt{2}\hspace{.5mm}\tilde{\mathbf{d}} \cos\omega t  \\
		\sqrt{2}\hspace{.5mm}\tilde{\mathbf{d}}\hspace{.5mm} \lambda \sin\omega t & \sqrt{2}\hspace{.5mm}\tilde{\mathbf{d}} \cos\omega t  &-2\hspace{.5mm}\tilde{\mathbf{d}}^2\hspace{1mm} \lambda \sin\omega t
	\end{pmatrix},
	\eea
	where
	\bea\label{lambda}
	\lambda= \frac{m\omega}{g_s^2},\hspace{1cm}\tilde{\mathbf{d}}= g_s\hspace{.2mm} \mathbf{d}.
	\eea 
	After doing some calculations, one obtains  
	\bea\label{Kp}
	{K}_{-}=
	-\alpha\begin{pmatrix}
		\cos\omega t &-\frac{1}{ \lambda}\sin\omega t  & 0 \\
		- {\lambda}\sin\omega t & -\cos\omega t   & 0  \\
		0 & 0  & 0 
	\end{pmatrix},
	\eea
	and 
	\bea\label{Km}
	{K}_{+}=
	-\alpha\begin{pmatrix}
		-\cos\omega t &\frac{1}{ \lambda}\sin\omega t & \sqrt{2}\hspace{.5mm}\tilde{\mathbf{d}} \cos\omega t  \\
		{\lambda}\sin\omega t & \cos\omega t   & -\sqrt{2}\hspace{.5mm}\tilde{\mathbf{d}}\hspace{.5mm}\lambda \sin\omega t   \\
		0 & 0  & 0 
	\end{pmatrix},
	\eea
	where $K^a_b=\Delta^{ac}\,k_{cb}$ in which $[\hat{\xi}^{a}_{\pm},\hat{\xi}^{b}_{\pm}] =i\Delta^{a b}_{\pm}$, 
	this implies 
	\bea
	\Delta_+^{a b}=\Delta_-^{a b}=
	\begin{pmatrix}
		0 &1  & 0 \\
		-1& 0 & 0  \\
		0 & 0  & 0 
	\end{pmatrix},
	\eea
	From the above expressions and \eqref{TCTFD.4},  the matrices $U_{\pm}(1)$ become  as follows 
	\bea\label{Upm}
	&&U_{-}(1)=
	\begin{pmatrix}
		\cosh\alpha-\cos\omega t \sinh\alpha&\lambda^{-1} \sin\omega t \,\sinh\alpha & 0 \\
		\lambda \sin\omega t\, \sinh\alpha& \cosh\alpha+\cos\omega t\, \sinh\alpha& 0  \\
		0 & 0  & 1
	\end{pmatrix},
	\eea
	$$ U_{+}(1)=\small
	\begin{pmatrix}
		\cosh\alpha+\cos\omega t \,\sinh\alpha&-\lambda^{-1} \sin\omega t\, \sinh\alpha&\sqrt{2}\hspace{.5mm}\tilde{\mathbf{d}} (1-\cosh\alpha-\cos\omega t\,\sinh\alpha)\\
		-\lambda \sin\omega t\, \sinh\alpha &\cosh\alpha-\cos\omega t\, \sinh\alpha& \sqrt{2}\hspace{.5mm}\tilde{\mathbf{d}}\hspace{.5mm}\lambda \sin\omega t\, \sinh\alpha\\
		0&0  &1
	\end{pmatrix}
	$$
	%with \bea && u_{11(22)}= \cosh(\alpha)\mp\cos(\omega t) \sinh(\alpha) , \hspace{1cm}u_{21}=\lambda^2 u_{12} = \lambda \sin(\omega t) \sinh(\alpha), \cr\nonumber\\&&u_{13}=\sqrt{2}\hspace{.5mm}\tilde{\mathbf{d}}_{q} \bigg(1-\cosh(\alpha)-\cos(\omega t)\sinh(\alpha)\bigg),\hspace{.5cm}u_{23}=\sqrt{2}\hspace{.5mm}\tilde{\mathbf{d}}_{q}\hspace{.5mm}\lambda \sin(\omega t) \sinh(\alpha).\nonumber\\\eea
	The equation (\ref{GsG0}) together with expressions for matrices $U_{\pm}(1)$ in (\ref{Upm}) and noting that 
	\bea
	G_{0,-}=
	\begin{pmatrix}
		\frac{1}{\lambda}&0 & 0 \\
		0&\lambda& 0  \\
		0 & 0  &2
	\end{pmatrix},\hspace{1cm}\hspace{2mm}G_{0,+}=
	\begin{pmatrix}
		4\hspace{.5mm}\tilde{\mathbf{d}}^2+\frac{1}{\lambda}&0 &2\sqrt{2}\hspace{.5mm}\tilde{\mathbf{d}} \\
		0&\lambda& 0  \\
		2\sqrt{2}\hspace{.5mm}\tilde{\mathbf{d}} & 0  &2
	\end{pmatrix}
	\eea
	gives us the two-point function of the target state as follows
	\bea \label{GTpm}
	\hspace{-.5cm}&&{G}_{\text{T},-}=
	\begin{pmatrix}
		g_{11,-}& g_{12,-}& 0 \\
		g_{21,-}&g_{22,-}  & 0  \\
		0 & 0  & 2
	\end{pmatrix},\hspace{1cm}
	{G}_{\text{T},+}=
	\begin{pmatrix}
		g_{11,+} & g_{12,+}&2\sqrt{2}\hspace{.5mm}\tilde{\mathbf{d}} \\
		g_{21,+}&  g_{22,+}& 0  \\
		2\sqrt{2}\hspace{.5mm}\tilde{\mathbf{d}} & 0  & 2
	\end{pmatrix},
	\eea
	where we have
	\bea
	&&g_{11,-}= \frac{1}{\lambda}( \cosh2\alpha-\cos\omega t\, \sinh2\alpha),
	\cr\nonumber\\
	&&g_{11_,+}=\frac{1}{\lambda}( 4\hspace{.5mm}\tilde{\mathbf{d}}^2\hspace{.5mm}{\lambda}+\cosh2\alpha+\cos\omega t\, \sinh2\alpha),
	\cr \nonumber\\
	&&g_{22,\pm}={\lambda}( \cosh2\alpha\mp\cos\omega t\, \sinh2\alpha),
	\cr \nonumber\\
	&&g_{12,-}=g_{21,-}=-g_{12,+}=-g_{21,+}=\sin\omega t\, \sinh2\alpha .
	\eea
	At this stage, to build the relative covariance matrix (\ref{Delta}), we need to find  the covariance matrix $G_R$ related to the reference state, which is found as  
	\bea\label{GRpm} 
	G_{R,+}=G_{R,-}=
	\begin{pmatrix}
		\frac{1}{\lambda_{R}}&0 & 0 \\
		0&\lambda_{R}& 0  \\
		0 & 0  & 2
	\end{pmatrix},
	\eea
	where 
	\bea\label{lambdaR}
	\lambda_{R} = m\omega_{R}/g_s^2.
	\eea
	Making use of (\ref{GTpm}) and (\ref{GRpm}), the relative covariance matrices $	{\cal M}_{\pm}$ become
	\bea\label{Deltapm} 
	{\cal M}_{-}=
	\begin{pmatrix}
		{\cal M}_{11,-}&	{\cal M}_{12,-} & 0 \\
		{\cal M}_{21,-} &	{\cal M}_{22,-} & 0  \\
		0 & 0  & 1
	\end{pmatrix},\hspace{1.5cm}
	{\cal M}_{+}=
	\begin{pmatrix}
		{\cal M}_{11,+} & 	{\cal M}_{12,+} &\sqrt{2}\hspace{.5mm}\tilde{\mathbf{d}} \\
		{\cal M}_{21,+} &	{\cal M}_{22,+}  & 0  \\
		2\sqrt{2}\hspace{.5mm}\tilde{\mathbf{d}}\hspace{1mm}\lambda_{R} & 0  &1
	\end{pmatrix},
	\eea
	where
	\bea
	&&	{\cal M}_{11_-}= \frac{\lambda_R}{\lambda}\Big( \cosh2\alpha-\cos\omega t\, \sinh2\alpha\Big),
	\cr\nonumber\\
	&&	{\cal M}_{11,+}= \frac{\lambda_R}{\lambda}\Big( 4\hspace{.5mm}\tilde{\mathbf{d}}^2\hspace{.5mm}{\lambda}+\cosh2\alpha+\cos\omega t\, \sinh2\alpha\Big),
	\cr\nonumber\\
	&&	{\cal M}_{22,\pm}=\frac{\lambda}{\lambda_R}\Big(\cosh2\alpha\mp\cos\omega t\, \sinh2\alpha\Big),
	\cr \nonumber\\&&
	{\cal M}_{21_+}= \lambda^2_{R} \hspace{1mm}	{\cal M}_{12_+} = -\lambda_R\hspace{.5mm}\sin\omega t\, \sinh2\alpha, 
	\cr \nonumber\\
	&&	{\cal M}_{21_-}=\lambda^2_R\hspace{.5mm}	{\cal M}_{12_-}= \lambda_R\sin\omega t\, \sinh2\alpha.
	\eea
	Finally, we use (\ref{ck2}), to compute the complexity of TFD state (\ref{TCTFD.3}) which can be written as follows 
	\bea\label{ck2app1}
	\mathcal{C}(t)=\frac{1}{4}\sum_{i=1}^{3}\bigg[\left(\log	{\cal M}^{(i)}_{+}\right)^2\hspace{.5mm}+\left(\log	{\cal M}^{(i)}_{-}\right)^2\bigg],
	\eea
	the eigenvalues of $	{\cal M}_{\pm}$ are denoted by $	{\cal M}^{(i)}_{\pm}$ which are given by 
	\bea\label{EigenDeltamm}
	{\cal M}^{(1)}_{-}=1,\hspace{1cm}	{\cal M}^{(2)}_{-}=A-\sqrt{A^2-1},\hspace{1cm}	{\cal M}^{(3)}_{-}=A+\sqrt{A^2-1},
	\eea
	where we have defined 
	\bea
	A\equiv\frac{1}{2\lambda \hspace{.5mm}\lambda_{R}}\bigg((\lambda^2+\lambda_R^2)\cosh2\alpha+(\lambda^2-\lambda_R^2)\cos\omega t\,\sinh2\alpha\bigg).
	\eea
	Note that the displacement $\tilde{\mathbf{d}}$  makes the computation of $	{\cal M}^{(i)}_{+}$ difficult, but to investigate the general behavior of $\mathcal{C} (t)$, in Appendix \ref{appC}, we expand  $	{\cal M}^{(i)}_{+}$ around $\tilde{\mathbf{d}}$ and present approximated expressions.  Even in the limit case of smaller $\lambda $ and $\tilde{\mathbf{d}}$, the corresponding eigenvalues are found to be 
	\bea\label{EigenDeltaspm}
	&&	{\cal M}^{(1)}_{\pm}(t)=1,\hspace{1cm}
	{\cal M}^{(3)}_{-}(t) = \frac{\lambda_{R}}{\lambda}\Big(\cosh2\alpha-\cos\omega t\,\sinh2\alpha\Big),
	\cr\nonumber\\
	&&
	{\cal M}^{(2)}_{+}(t)= 4 \hspace{.5mm}\tilde{\mathbf{d}}^2\hspace{.5mm}\lambda_{R},\hspace{.5cm}
	{\cal M}^{(3)}_{+}(t) = 4 \hspace{.5mm}\tilde{\mathbf{d}}^2\hspace{.5mm}\lambda_{R} +\frac{\lambda_{R}}{\lambda}\Big(\cosh2\alpha+\cos\omega t\,\sinh2\alpha\Big).
	\eea
	\begin{figure}[H]
		\centering
		\includegraphics[scale=.23]{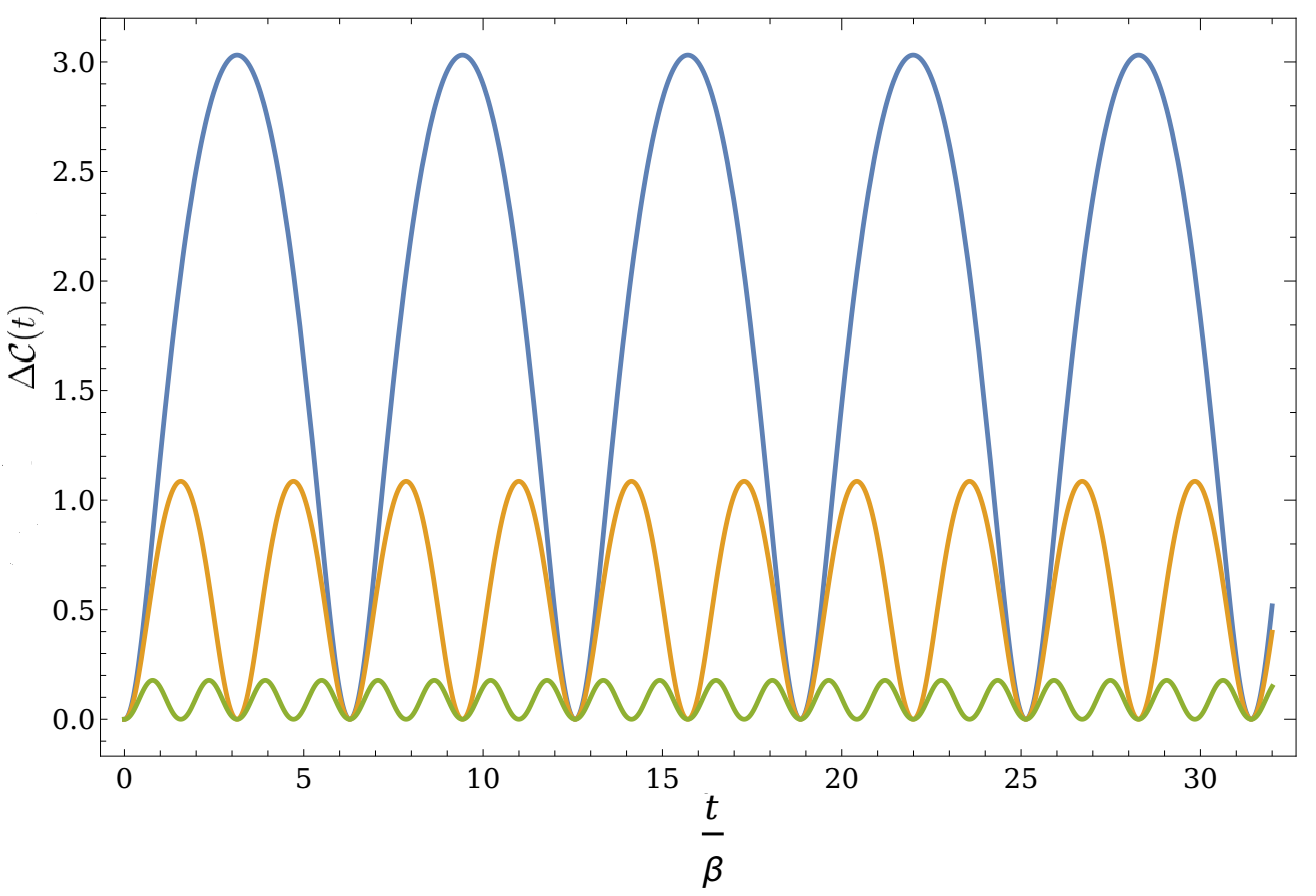}
		\includegraphics[scale=.23]{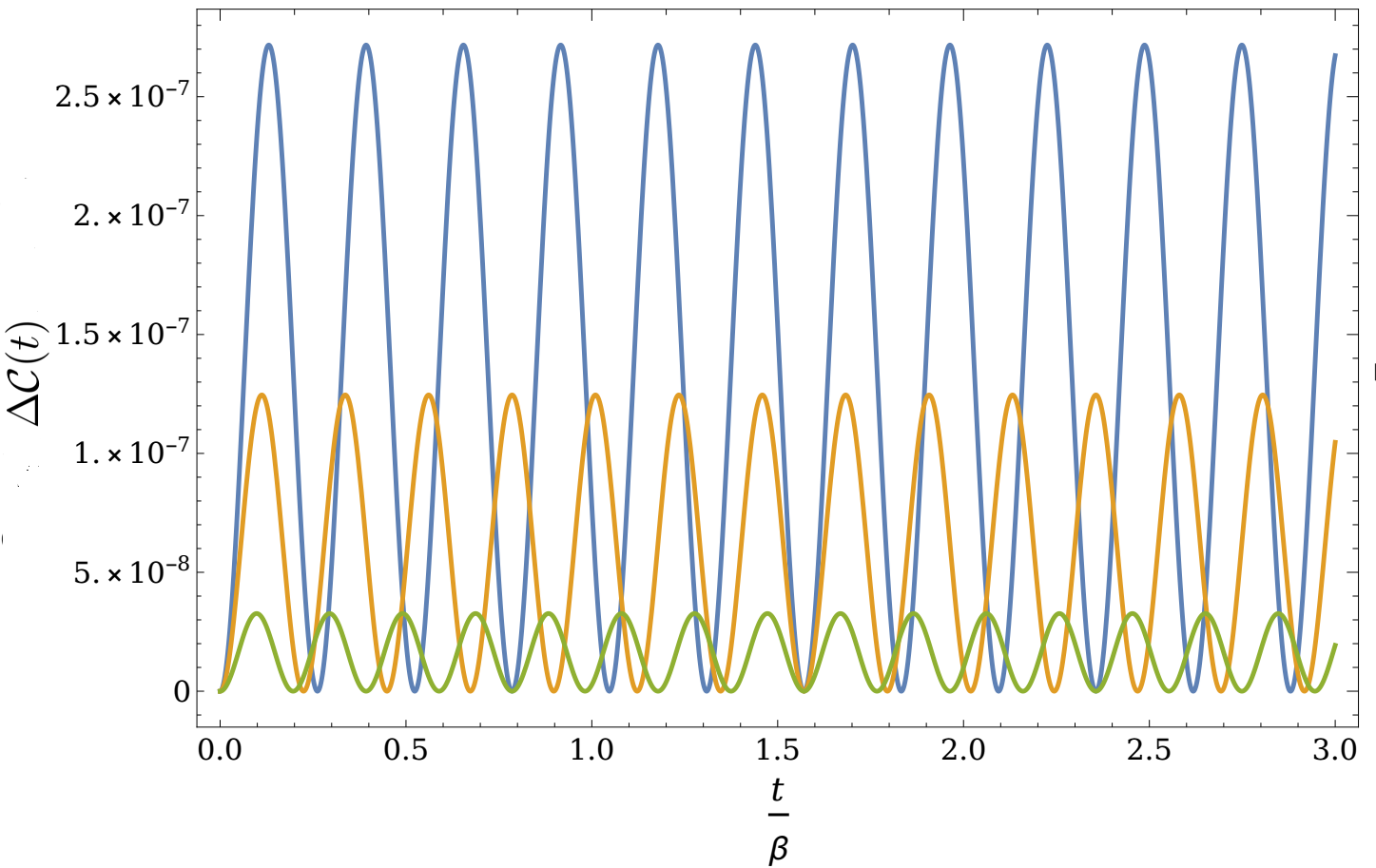}
		\caption{ The  complexity of TFD state ($\Delta\mathcal{C}(t)=\mathcal{C}(t)- \mathcal{C}(0)$) in the simple limit for  $\lambda_{R}=1$, $\beta \omega_{R} = 10$ and different values of $\beta \omega$.  Left panel: $\beta \omega=\frac{1}{2}$\hspace{.1mm}(blue), 1\hspace{.1mm}(yellow) and 2\hspace{.1mm}(green). Right panel: $\beta \omega=12$\hspace{.1mm}(blue), 14\hspace{.1mm}(yellow), and 16\hspace{.1mm}(green) As can be seen, in the absence of external factors, $\Delta\mathcal{C}(t)$  decreases significantly with increasing frequency.  }	\label{myerssimple}\label{fig1}
	\end{figure}
	\begin{figure}[htb!]
		\centering\includegraphics[scale=.2]{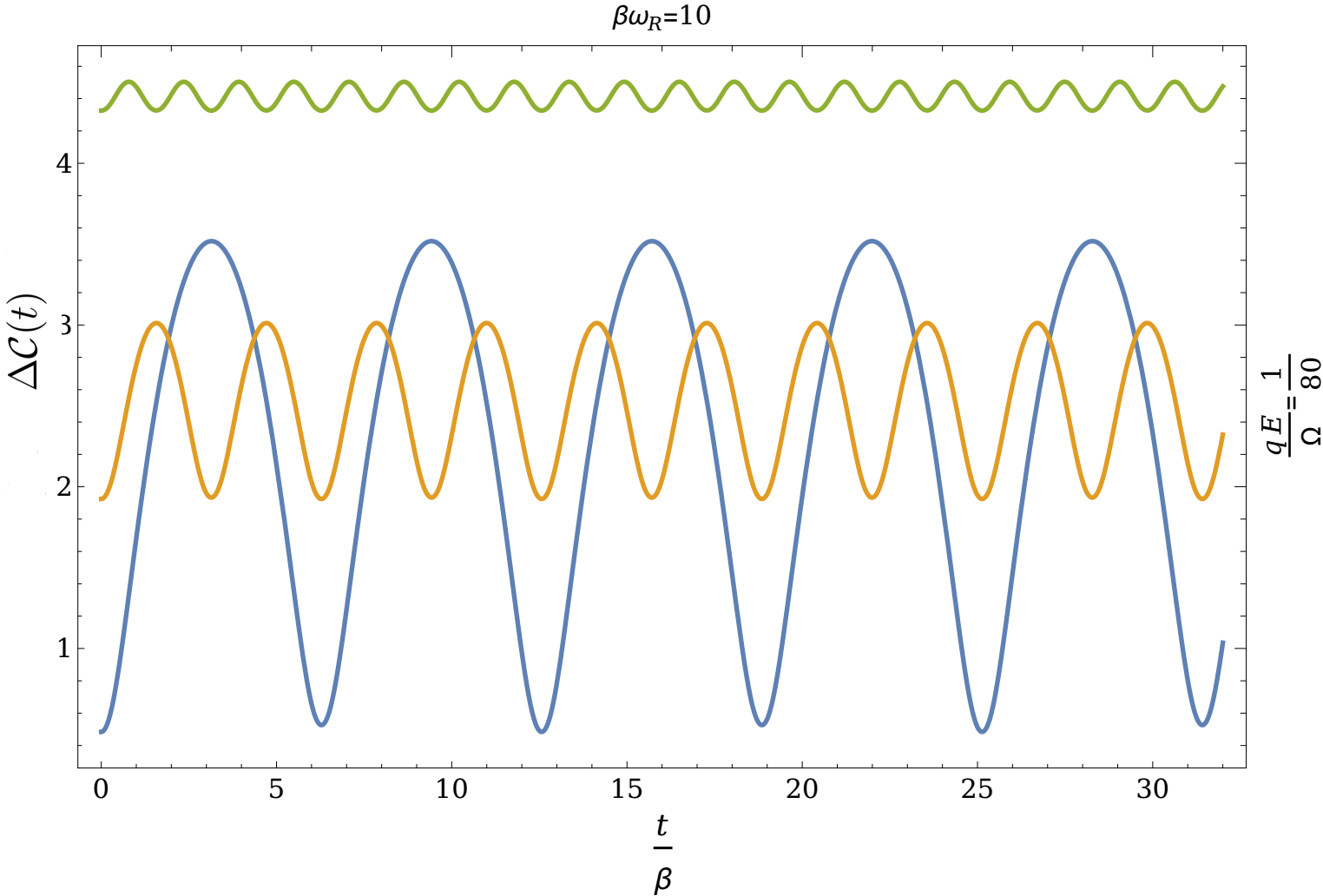}\hspace{.4cm}
		\includegraphics[scale=.2]{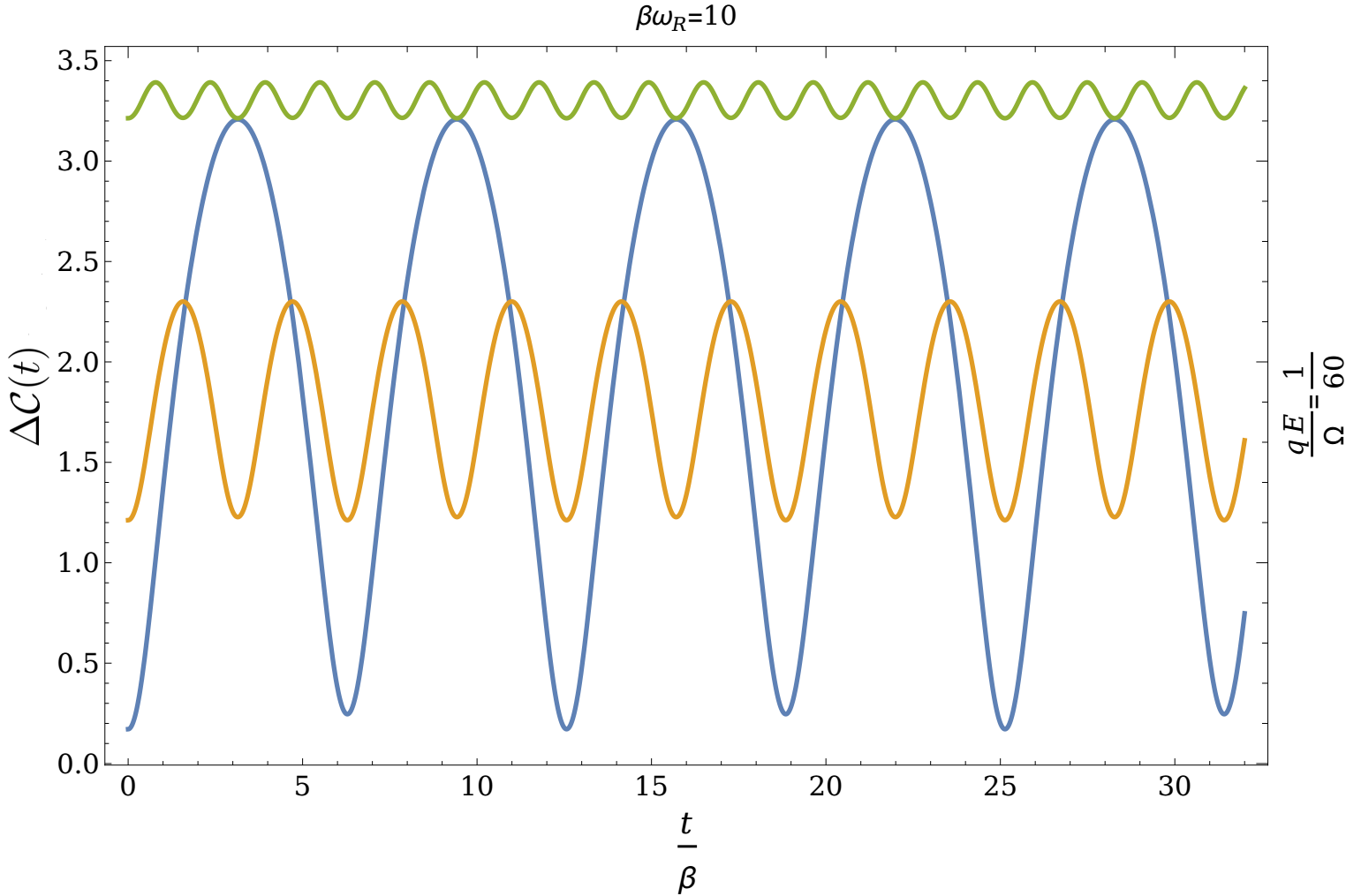}\vspace{.4cm}
		\includegraphics[scale=.2]{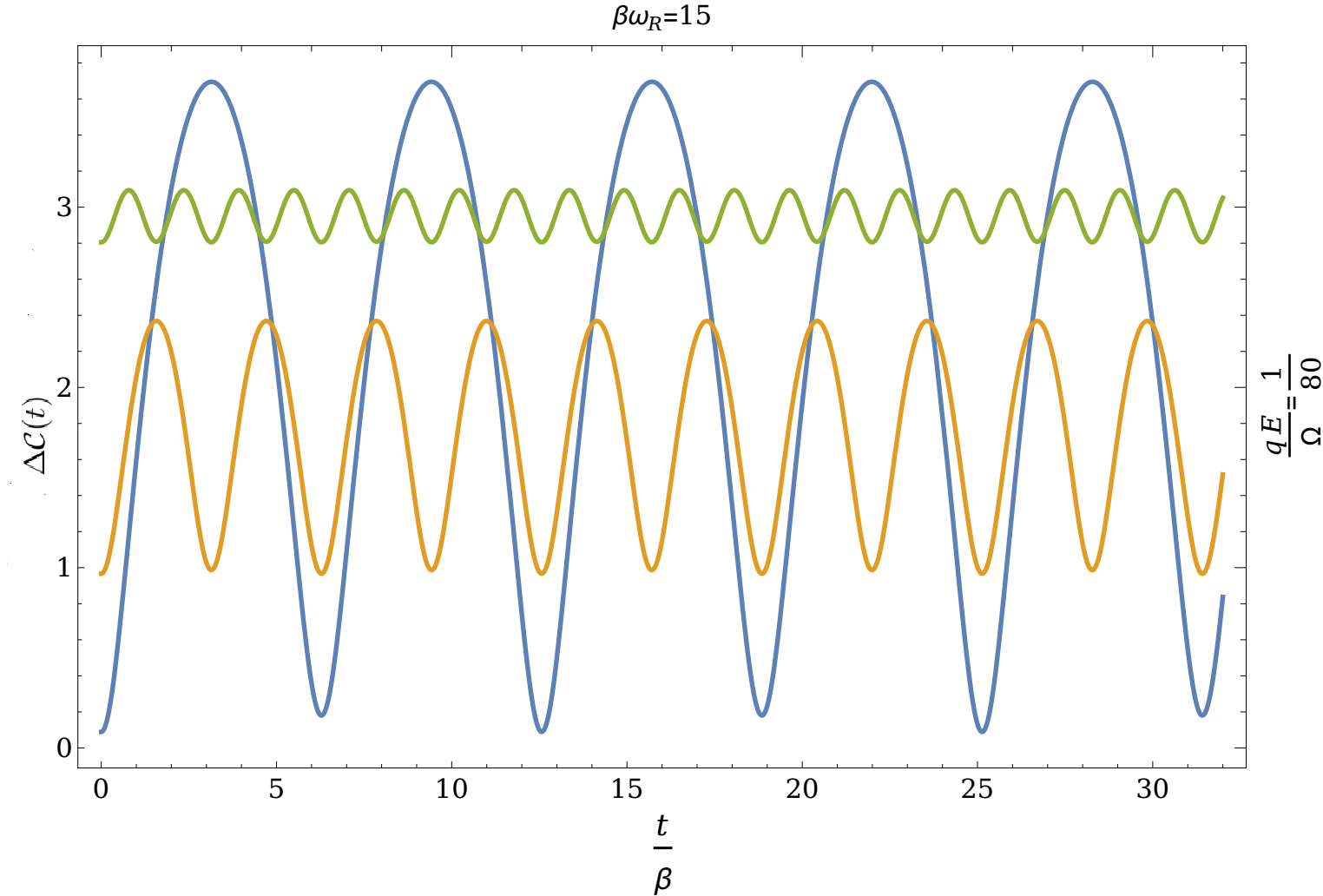}\hspace{.4cm}
		\includegraphics[scale=.31]{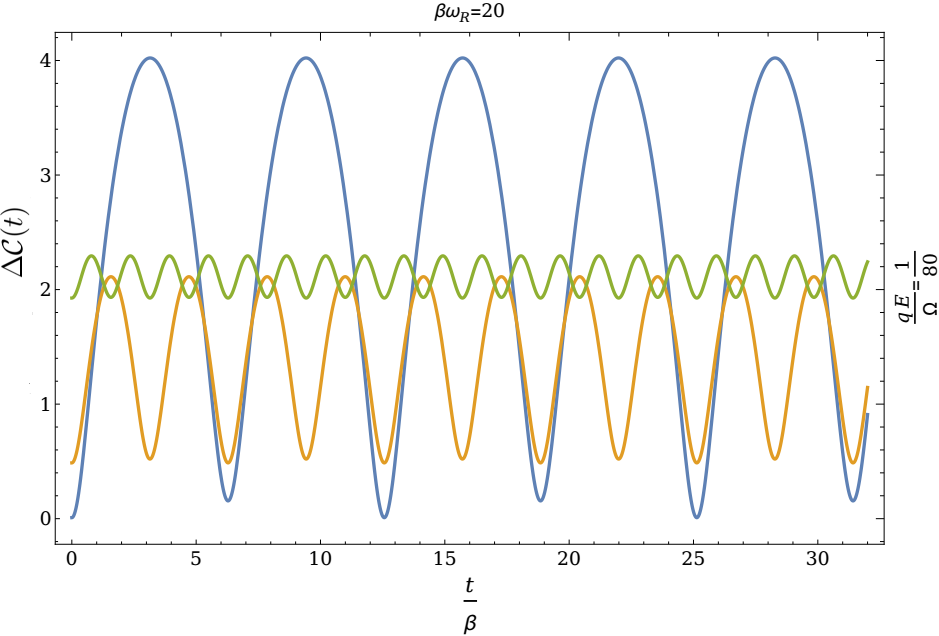}
		\caption{The temporal evolution of the complexity of the TFD state under the influence of a minute electric field with $\lambda_{R}=1$ is examined. Specifically, when subject to a very small electric field, the cases of $\beta \omega=\frac{1}{2}$ (blue), 1 (yellow), and 2 (green) are considered. The first line of the figure corresponds to $\beta \omega_{R}=10$, whereas in the subsequent line, the left panel pertains to $\beta \omega_{R}=15$, and the right panel corresponds to $\beta \omega_{R}=20$. Upon the introduction of a small yet non-negligible external field, it is observed that $\Delta\mathcal{C}(t)$ exhibits an increase with rising frequency. Interestingly, under identical conditions, the initial reference state frequency $\beta \omega_{R}$ acts as a governing factor, leading to a decrease in complexity as it is augmented.  }\label{fig2}
	\end{figure}	
	Now we have all the ingredients to study the complexity (\ref{ck2app1}) of the time-dependent state (\ref{TCTFD.3}). We try to present the results of the complexification in various cases, noting that we set $\lambda_{R}=1$ and the dimensionless ratio $q E/ \Omega$ has been defined where $\Omega= \omega g_s$. Moreover, we have defined $\Delta\mathcal{C}(t)$ as
	\bea\label{deltack2}
	\Delta\mathcal{C}(t)= \mathcal{C}(t)- \mathcal{C}(0).
	\eea
	In Figure \ref{fig1} we have shown the evolution of $\Delta\mathcal{C}(t)$  when the external electric field was turned off. It can be observed that the complexity of the states decreases significantly with increasing frequency. However, when the electric field is activated, the situation changes. In figure \ref{fig2},  we have studied the effect of a small but non-vanishing external electric field on the complexity; We observe that high-frequency systems have lower $\Delta\mathcal{C}(t)$ in the presence of a weak electric field, despite the previous case. We can also say that the frequency of the initial reference state  $\beta\omega_{R}$ plays the role of a regulator, so that in the same state the complexity decreases as it increases. In Figure \ref{fig4} we repeat the same numerical analysis for larger values of the external electrical field. Interestingly, the system returns to its original behavior as the external electric field increases; This means that the complexity of the system decreases as the frequency increases. In this regime, we also observe that $\Delta\mathcal{C}(t)$  gets a decreasing rate as the frequency of the reference system decreases.
	
	Here, let us end this section with a remark. As mentioned, to compute the complexity, one needs to identify a unitary operator that acts on the reference state to build the target state. There are some proposals say as the Nielsen geometric proposal (used in this paper) and the Margolus-Levitin and Lloyd methods \cite{Margolus:1997ih, 234}\footnote{ For a comprehensive review we refer to  \cite{Deffner:2017cxz}.}, which the latter determines a limit on quantum computation by considering the minimum time as a quantum state evolves to its orthogonal (distinguishable) states. It is shown that the rate of complexity is bounded by the orthogonality time as $\frac{d}{dt}{\mathcal{C} }\le\frac{1}{\tau_\bot}$, and this implies the time variation of the complexity is controlled by the system's energy and the maximal rate of complexity becomes $\frac{d}{dt}{\mathcal{C} }\le\frac{2\langle H\rangle}{\pi}$, (in units $\hbar=c=1$). In this article, although a different aspect of complexity for a harmonic oscillator was investigated, it is interesting to mention that based on Margolus-Levitin and Lloyd methods, in  Ref.s \cite{REZA:tanhayi, Khorasani:2021zus} the rate of complexity for a system consisting of a set of harmonic oscillators in the presence of an electric field was studied, and it was observed that frequency plays an important role in the rate of complexity. On the other hand, the effect of the external electric field causes the upper bound of the complexity rate to relax to a smaller value.

	%%%%%%%%%%%%%%%%%%%%%%%%%%%%%%%%%%%%%%%%%%%%%%%%%%%%%%

	\section{Conclusions}

	By definition, computational complexity is the minimum number of operations needed to obtain the target state from an initial reference state. Making use of the covariance matrix approach, for the TFD state,  the complexity of a charged TFD state in the presence of an electric field has been studied in Ref. \cite{Doroudiani:2019llj}. In this approach, by specifying a cost function, one should optimize every possible path starting from the reference state and leading to the construction of the final target state. For Gaussian modes, the relative covariance matrix, which is written based on the definition of the two-point functions related to the reference and target modes, should be constructed.  It is argued that this matrix could carry the information on the complexity of the state.  In this paper, we followed the same approach of the covariance matrix and Nielsen's geometric approximation and extended the results in parts to a harmonic oscillator. We first built $\Delta\mathcal{C}$ the complexity of state as a function of time and then investigated its behavior against the external electric field as well as the frequency of the system and reference state. An electric field induces a shift in the energy level, which can also be viewed as a displacement in the corresponding state due to the fixed gates employed to generate the shifted TFD state. This motivates the analysis of the coordinate dependence of Nielsen complexity. By introducing the displacement in the state, we defined the corresponding gate operator and demonstrated its significance for the circuit design (Appendix \ref{appB}). Furthermore, we examined the Nielsen complexity and found that the complexity of the state dropped sharply as the system’s frequency increased when the external field was switched off. However, the influence of the external electric field on the complexity is not trivial, with a small value of the external field the complexity increases with increasing frequency, but with a larger value of the electric field, the system returns to its previous behavior without an electric field, which means that $\Delta \mathcal{C} (t)$ decreases with increasing frequency. Moreover, our results showed that the frequency of the reference state also affects the complexity as by increasing  $\beta\omega_R$ while keeping other parameters unchanged, the complexity decreases with its increase. \\ As a comment, to consider the effect of changing the reference state by rescaling $g_s$, one can say that as long as in most cases, we are dealing with the ratio of  $\frac{\lambda_{R}}{\lambda}$, thus any rescaling of $\lambda_{R}$ (by $g_s$) is compensated in ratio $\frac{\lambda_{R}}{\lambda}$ by the same rescaling of $\lambda$ (equations \eqref{lambda} and \eqref{lambdaR}). In this sense, the complexity becomes independent of the rescaling of the reference state. However, in the presence of the electric field, the complexity apart the combination $\frac{\lambda_R}{\lambda}$ also depends on $\tilde{\mathbf{d}} \lambda_{R}^2$ and $\tilde{\mathbf{d}} \lambda \lambda_{R}$ combinations. These combinations change by rescaling the $\lambda_{R}$, consequently, the complexity would also change.\\ For future work, it would be fascinating to investigate the effect of supposing non-trivial gate scale $(\lambda_R\neq1)$ and also external factors on the complexity and its dynamics for the Gaussian states of the harmonic oscillators.

	\begin{figure}[H]
		
		\includegraphics[scale=.22]{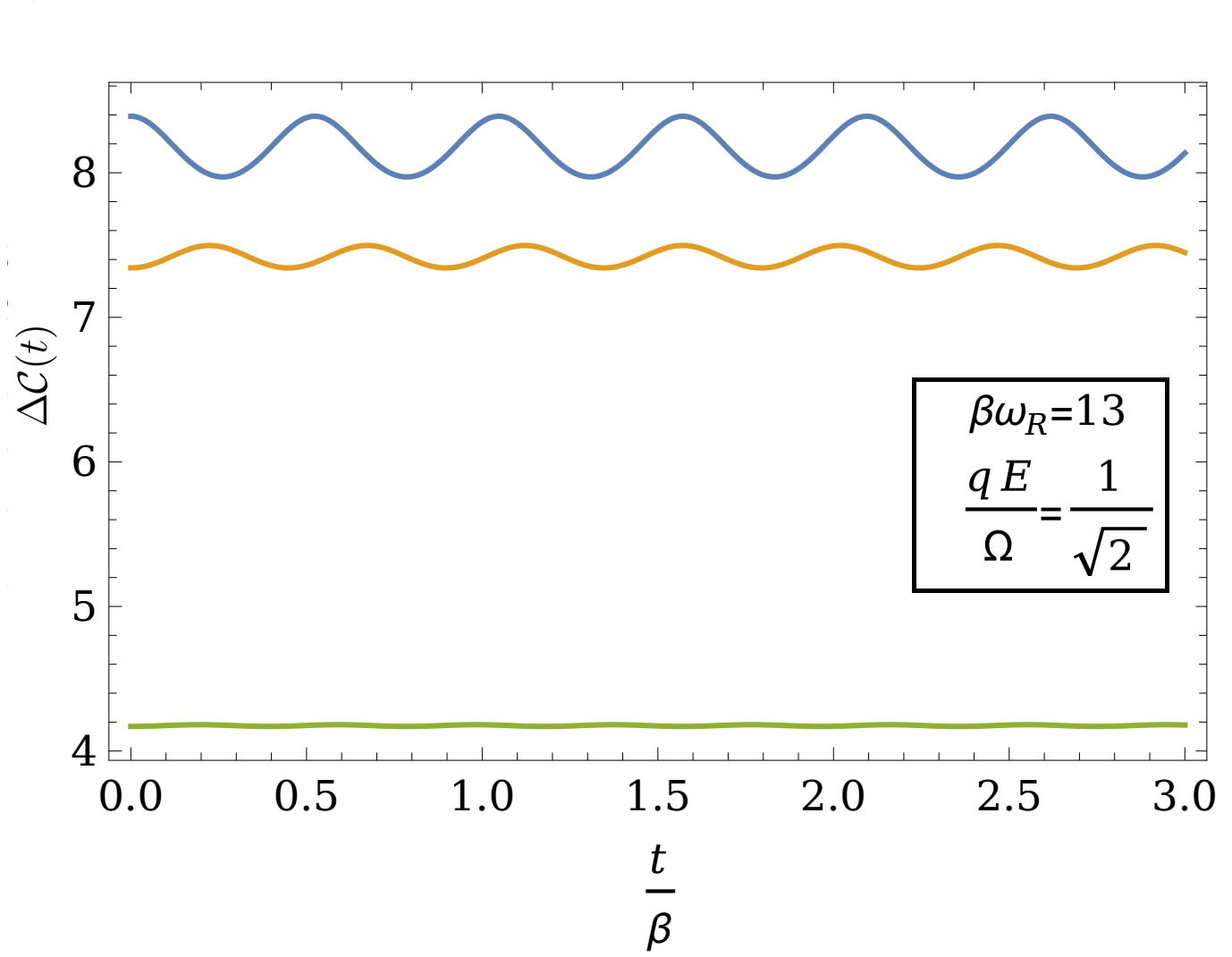}\hspace{.4cm}
		\includegraphics[scale=.23]{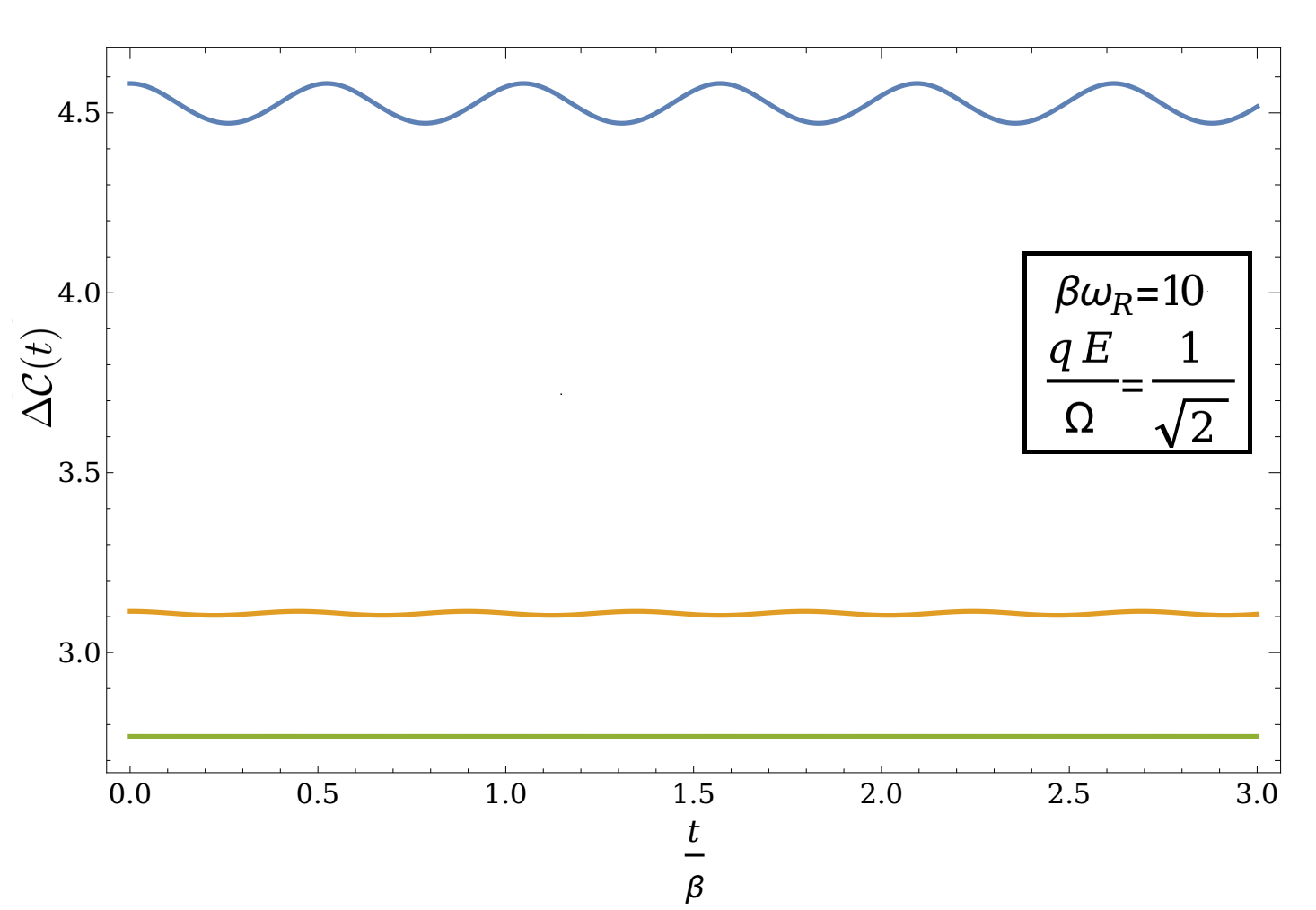}\vspace{.5cm}
		\includegraphics[scale=.23]{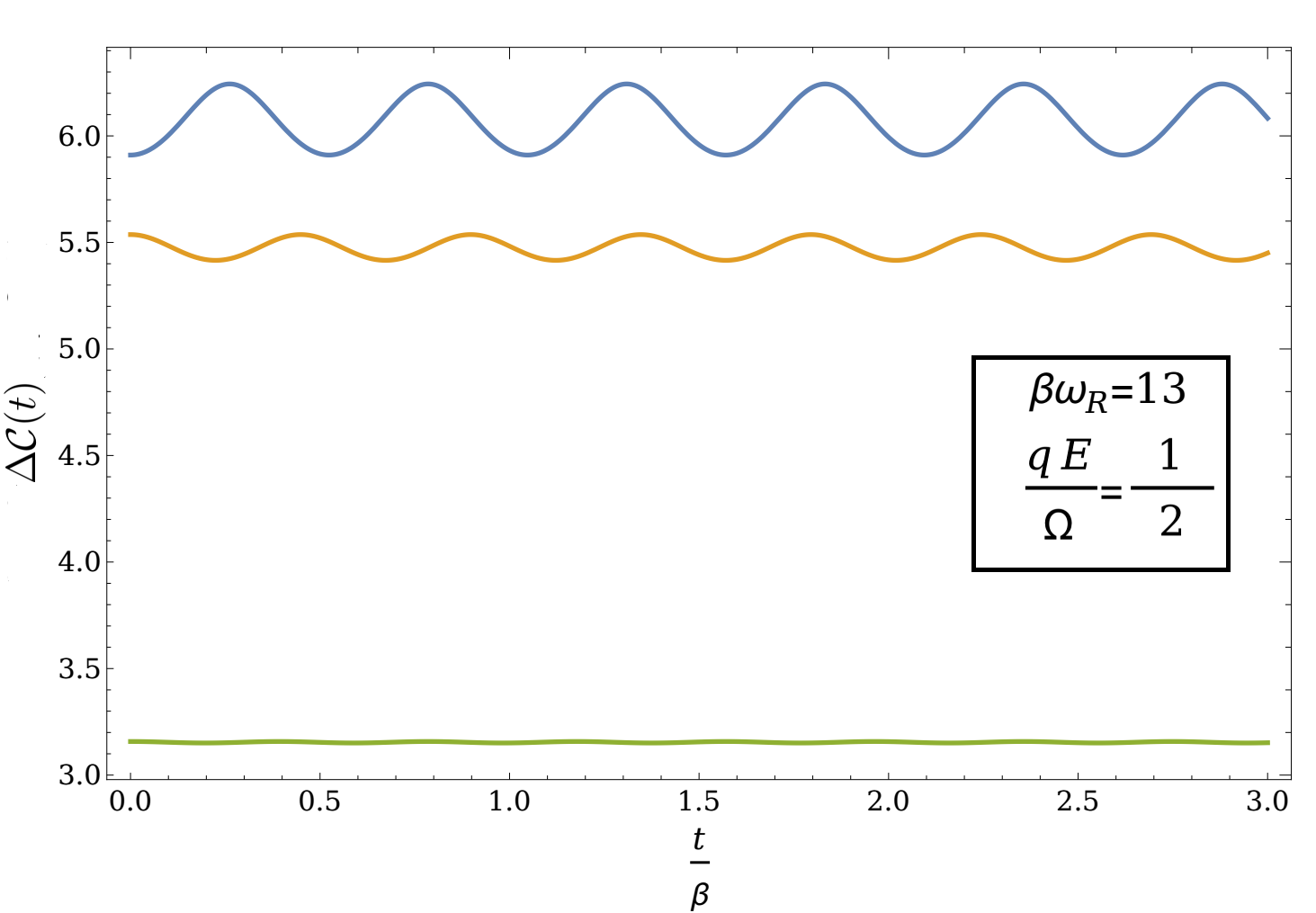}
		\includegraphics[scale=.23]{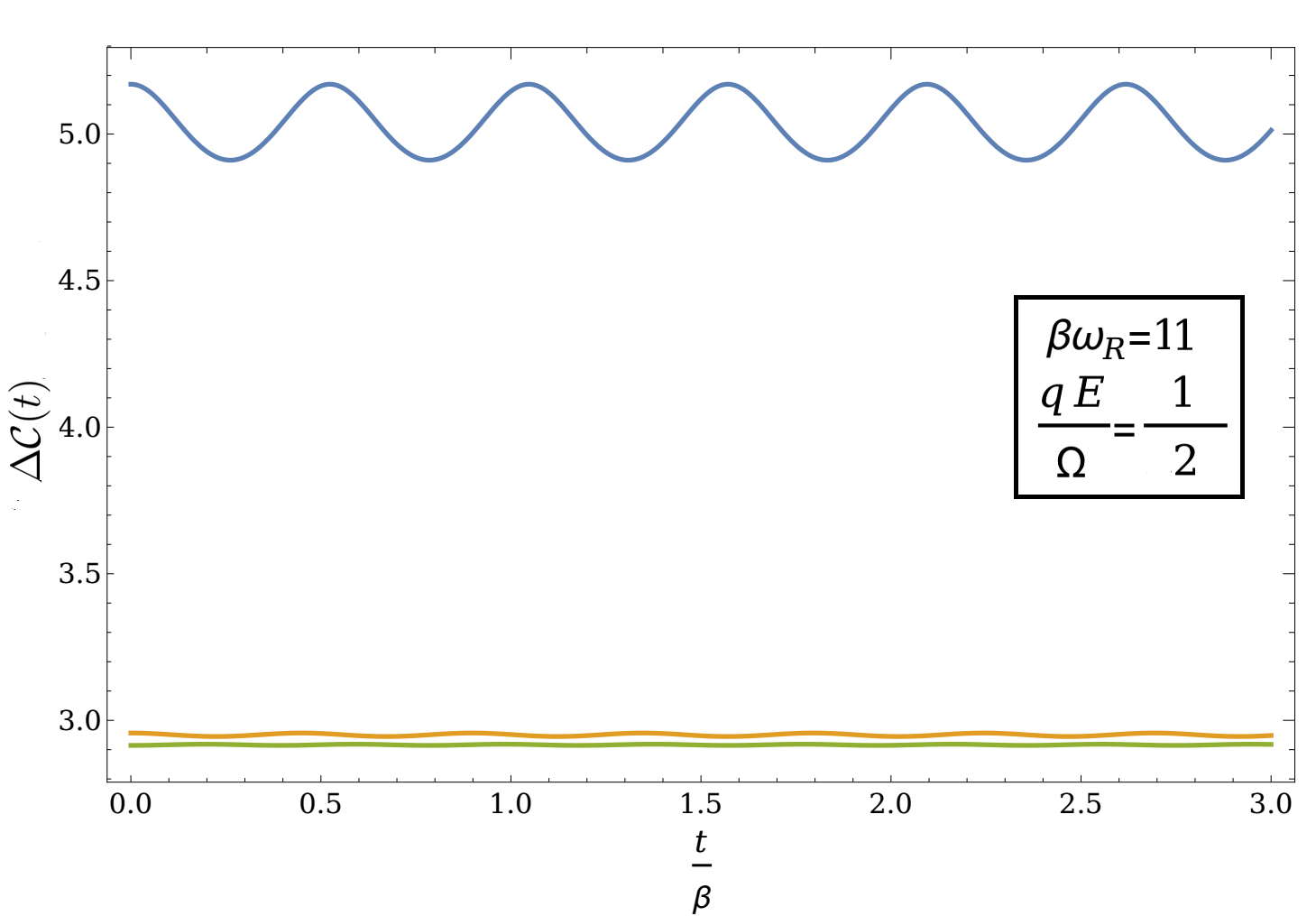}\hspace{.6cm}
		\caption{ The variation of the complexity of TFD state as a function of time for $\lambda_{R}=1$  with an external electric field. We set $\beta \omega=\frac{1}{2}$\hspace{.1mm}(blue), 1\hspace{.1mm}(yellow), and 2\hspace{.1mm}(green). First line: we consider $\frac{qE}{\omega}\sim 0.7$ and $\beta\omega_{R}=13$ and 10. Second line: $\frac{qE}{\omega}=0.5$ is supposed. In this case, for larger electric field values, the system behaves like its initial behavior, namely,  $\Delta\mathcal{C}(t)$ decreases as the frequency increases, one also observes that it gets a decreasing rate as the frequency of the reference system decreases.}\label{fig4}
	\end{figure}

	\subsection*{Acknowledgments}
	
	We would like to thank M. Alishahiha for useful comments.  R. Pirmoradian would also like to thank A. Naseh for his very helpful comments and related discussions. R. Pirmoradian would also like to extend special thanks to  M. Reza Lahooti Eshkevari dean of Ershad Institute and M. Saidi for their support. We would like to thank T. Hamedi for her comments. We would like to thank the anonymous referee for his/her valuable comments and suggestions, which have improved the quality and clarity of the paper.

	%%%%%%%%%%%%%%%%%%%%%%%%%%%%%%%%%%%%%%%%%%%%%%%
	%--------------------------------------------------------------------------
	\appendix

	\section{
		Decomposition of TFD in diagonal basis}\label{appA}
	Our goal is to construct a state by coupling two harmonic oscillators that are labeled as $L$ and $R$ and one can show that 
	\bea
	[\hat{a}_{L}\hspace{.5mm}\hat{a}_{R},\,\hat{a}^\dagger_{L}\hspace{.5mm}\hat{a}^\dagger_{R} ]=\hat{a}^\dagger_{L}\hspace{.5mm}\hat{a}_{L}+\hat{a}^\dagger_{R}\hspace{.5mm}\hat{a}_{R}+1.
	\eea
	By defining $\hat K_\pm$ and $\hat K_0$ as
	$$\hat K_+\equiv\hat{a}^\dagger_{L}\hspace{.5mm}\hat{a}^\dagger_{R},\,\,\,\,\hat K_-\equiv\hat{a}_{L}\hspace{.5mm}\hat{a}_{R},\,\,\,\,\,\hat K_0\equiv\frac{1}{2}\Big(\hat{a}^\dagger_{L}\hspace{.5mm}\hat{a}_{L}+\hat{a}^\dagger_{R}\hspace{.5mm}\hat{a}_{R}+1\Big)$$
	it is easy to verify that they satisfy the following commutation relations
	\bea
	[\hat K_-,\hat K_+]=2\hat K_0,\,\,\,\,[\hat K_0,\hat K_+]=\hat K_+,\,\,\,\,[\hat K_0,\hat K_-]=-\hat K_-\,,
	\eea
	one may assume that they are the elements of the $su(1,1)$ group. Ref. \cite{boo} has studied the decomposition of the element for this group in detail, and in our case, the decomposition is given by 
	\bea
	e^{\alpha(\hat{K}_+-\hat{K}_-)}=e^{(\tanh\alpha)\hat{K}_+}\hspace{1mm}e^{-2(\log\cosh\alpha)\hat{ K}_0}\hspace{1mm}e^{-(\tanh\alpha)\hat{K}_-}.
	\eea
	On the other hand, one can write
	\bea\label{OO}
	e^{\alpha(\hat{K}_+-\hat{K}_-)}\mid\hspace{-1mm}0\rangle_L\hspace{.5mm}\hspace{-.5mm}\mid \hspace{-1mm}0\rangle_R=\frac{1}{\cosh\alpha}e^{(\tanh\alpha)\hat{a}^\dagger_{L}\hspace{.5mm}\hat{a}^\dagger_{R}}\mid\hspace{-1mm}0\rangle_L\hspace{-.5mm}\mid \hspace{-1mm}0\rangle_R,
	\eea
	where we have used the following relations
	\bea
	{K}_- \mid\hspace{-1mm}0\rangle_L\hspace{.5mm}\mid\hspace{-1mm} 0\rangle_R=0,\hspace{1cm} \hspace{3mm}{K}_0\mid\hspace{-1mm}0\rangle_L\hspace{.5mm}\mid\hspace{-1mm} 0\rangle_R=\frac{1}{2}\mid\hspace{-1mm}0\rangle_L\mid\hspace{-1mm} 0\rangle_R.
	\eea 
	Back to the TFD state  (\ref{TFD.2}), it can be written as
	\bea\label{TFD.3}
	\sqrt{1-e^{-\beta\omega}}\hspace{1mm}\exp{\left(e^{-\frac{\beta \omega}{2}}\hspace{1mm}\hat{a}^\dagger_{L}\hat{a}^\dagger_{R}\right)}\mid\hspace{-1mm}0\rangle_L\hspace{.5mm}\mid \hspace{-1mm}0\rangle_R&=&\frac{1}{\cosh\alpha}e^{(\tanh\alpha)\hat{a}^\dagger_{L}\hspace{.5mm}\hat{a}^\dagger_{R}}\mid\hspace{-1mm}0\rangle_L\hspace{-.5mm}\mid \hspace{-1mm}0\rangle_R\nonumber\\
	&=&e^{\alpha\left(\hat{a}^\dagger_{L}\hspace{.5mm}\hat{a}^\dagger_{R}-\hat{a}_{L}\hspace{.5mm}\hat{a}_{R}\right)}\mid\hspace{-1mm}0\rangle_L\hspace{.5mm}\mid\hspace{-1mm} 0\rangle_R,
	\eea
	where we have used \eqref{OO} and $\cosh \alpha=\frac{1}{\sqrt{1-e^{-\beta\omega}}}$. 
	To define, the target state and track the position-dependency of the complexity,  in the basis given by \eqref{xpm},  the state (\ref{TFD.3}) turns to 
	\bea
	\mid \hspace{-1mm}\Psi_T\rangle= e^{-i\alpha(\hat{x}_+\hat{p}_+-\sqrt{2}\mathbf{d} \hat{p}_+)}\mid\hspace{-1mm}0\rangle_+\otimes e^{i\alpha\hat{x}_-\hat{p}_-}\hspace{-1mm}\mid\hspace{-1mm}0\rangle_-,
	\eea
	where in writing the above equation and also equation \eqref{TCTFD.3}, the following relations have been used 
	\bea
	\hat{a}^\dagger_{L}\hat{a}^\dagger_{R}-\hat{a}_{L}\hat{a}_{R}
	&=& - i \left(\hat{x}_L \,\hat{p}_R+\hat{p}_L\,\hat{x}_R \right)+i\mathbf{d}(\hat{p}_L+\hat{p}_R)\nonumber\\
	&=&- i \left(\hat{x}_+ \hat{p}_+ -\hat{x}_- \hat{p}_-\right)+i\sqrt{2}\mathbf{d} \hat{p}_+\nonumber\\
	\hat{a}^\dagger_{L}\hat{a}^\dagger_{R}+\hat{a}_{L}\hat{a}_{R}
	&=& {m\omega}\Big(\hat x_L\,\hat x_R-\mathbf{d} (\hat x_L+\hat x_R)+\mathbf{d} ^2-\frac{\hat p_L\,\hat p_R}{m^2\omega^2}\Big)\nonumber\\
	&=&\frac{m\omega}{2}\Big(\hat x_+^2-\hat x_-^2-2\sqrt{2}\mathbf{d}\hat x_++2\mathbf{d}^2-\frac{\hat p_+^2}{m^2\omega^2}+\frac{\hat p_-^2}{m^2\omega^2}\Big). 
	\eea
	Let us also comment on the corresponding wave function. In the normal coordinate, the reference state is simply defined by 
	\be \Psi_R(x_\pm)=(\frac{m\omega_R}{\pi})^\frac{1}{4}\,e^{-\frac{m\omega_R}{2}x_\pm^2},\label{ref}
	\ee
	 and we define the ground states as
	 \bea
	 \Psi_0( x_-)&=&(\frac{m\omega}{\pi})^\frac{1}{4}\,\exp\Big[-\frac{m\omega}{2}x_-^2\Big],\nonumber\\
	 \Psi_0( x_+)&=&(\frac{m\omega}{\pi})^\frac{1}{4}\,\exp\Big[-\frac{m\omega}{2}(x_+-\sqrt{2}\mathbf{d})^2\Big].
	 \eea
The corresponding target state is also followed by 	   
	\bea\label{a1}
	\Psi( x_-)&=&	\langle x_{-}\hspace{-1mm}\mid \hspace{-1mm}e^{i\alpha {x}_-{p}_-}\hspace{-1mm}\mid\hspace{-1mm}0\rangle_-\nonumber\\
	&=&	(e^{\alpha\,x_-\frac{d}{dx_-}})\langle x_{-} \hspace{-1mm}\mid0\rangle_-\simeq (1+\alpha\,x_-\frac{d}{dx_-})\langle x_{-} \hspace{-1mm}\mid0\rangle_-\nonumber\\
	&\simeq&	\Psi( e^\alpha\,x_-)	=(\frac{m\omega}{\pi})^\frac{1}{4}\,\exp\Big[-\frac{m\omega}{2}e^{2\alpha}x_-^2\Big],
	\eea
	and similarly for $x_+$, one obtains
	\bea\label{a2}
	\Psi( x_+)=	\langle x_{+}\hspace{-1mm}\mid \hspace{-1mm}e^{-i\alpha(\hat{x}_+\hat{p}_+-\sqrt{2}\mathbf{d} \hat{p}_+)}\hspace{-1mm}\mid\hspace{-1mm}0\rangle_+=(\frac{m\omega}{\pi})^\frac{1}{4}\,\exp\Big[-\frac{m\omega}{2}e^{-2\alpha}(x_+-\sqrt{2}\mathbf{d})^2\Big],
	\eea
	consequently, one can write
	\bea\label{tar} \Psi_T(x_-,x_+)=\sqrt{\frac{m\omega}{\pi}}\,e^{(-m\omega \mathbf{d}^2\,e^{-2\alpha}) }\exp\Big[-\frac{m\omega}{2}e^{2\alpha}\,x_-^2-\frac{m\omega}{2}e^{-2\alpha}\Big(x_+^2-2\sqrt{2}\mathbf{d}\,x_+\Big)\Big],
	\eea
	similarly, one can use \eqref{TCTFD.3} to write the time-dependent wave function.

	%-----------------------------------------------
	%--------------------------------------------------------------------------
	\section{Circuit for a toy model}\label{appB}
	In this appendix, we examine the position-dependency of a circuit in a toy model. In principle, the $\hat U$ operator transforms the reference state to the target state as 
	\bea\
	\mid\hspace{-1mm}\Psi_{T}\rangle=\hat{ U} \mid\hspace{-1mm}\Psi_{R}\rangle,
	\eea
	could be constructed by the string of $n$-gates, as 
	$$\hat U={g_n}\,g_{n-1}...g_2\,g_1.$$
	A circuit is a configuration of gates that produces the desired target state for the $\hat U$ operator. The circuit depth is the total number of gates in a configuration, and the circuit or gate complexity is the minimum number of gates required for a circuit.\footnote{ It should be noted that the circuit complexity of states is determined by the minimal circuit that prepares the state from a reference state. However, the circuit complexity of unitaries reflects the system size and is quantified by the gate count of the minimal circuit that implements the unitary. In our specific model, these complexities are equivalent.} Let us define a toy model where the reference and target states are given by ($m_-=m_+=1$)
	\bea \label{tar123} \Psi_R(x_-,x_+)&=&(\frac{\omega_R}{\pi})^{\frac{1}{2}}\,\exp\Big[-\frac{\omega_R}{2}(x_-^2+x_+^2)\Big],\\\label{tar1} \Psi_T(x_-,x_+)&=&(\frac{\omega_-\omega_+}{\pi^2})^{\frac{1}{4}}\,e^{(-\omega_+\mathbf{d}^2e^{-2\alpha}) }\exp\Big[-\frac{\omega_-}{2}\,e^{2\alpha}\,x_-^2-\frac{\omega_+}{2}\,e^{-2\alpha}\,x_+^2+\sqrt{2}\omega_+\,e^{-2\alpha}\,\mathbf{d}\,x_+\Big],\nonumber
	\eea
	we want to construct the $\hat U$ operator. Based on quantum mechanics, the general operators that are used to construct the gates are $\hat x_-$ and $ \hat x_+$ operators and their momenta $\hat p_-\!=\!-i\partial_-,\,\hat p_+\!=\!-i\partial_+$. In our case to construct a circuit, we shall explore the role of the operator $\mathbf{d}$ where based on, the displacement operator/gate can be defined as $e^{i\epsilon\mathbf{d}\hat{p}}$. Therefore, a set of elementary gates can be defined as 
	\bea	g_{ab}=e^{i\epsilon \hat x_a\hat p_b},\, \,\,\,\, g_{aa}=e^{i\epsilon\frac{ \hat x_a\hat p_a+\hat p_a\hat x_a}{2}}~,\,\,\,\,\,g_{\mathbf{d}a}=e^{i\epsilon \mathbf{d} \hat p_a}~,\,\,\,\,\,g_{a0}=e^{i\epsilon \hat x_a p_0},
	\eea
	where $a$ and $b$ run over $\{-,+\}$ and $x_0$,  $p_0$ are constant.  One can check that 
	\bea
	g_{ab}\,\Psi(x_a,x_b)&=&\Psi(x_a\,,x_b+\epsilon x_a),\nonumber\\
	g_{aa}\,\Psi(x_a,x_b)&=&e^{\epsilon/2}\Psi(e^\epsilon x_a\,,x_b),\nonumber\\
	g_{\mathbf{d}a}\,\Psi(x_a,x_b)&=& \Psi( x_a+\epsilon\mathbf{d} \,,x_b),\nonumber\\
	g_{a0} \,\Psi(x_a,x_b)&=&e^{i\epsilon p_0x_a}\,\Psi(x_a,x_b).
	\eea
	In this case, we are interested in finding the $\hat{ U}$ operator that be defined by 
	\begin{equation}
		\hat U= (g_{\mathbf{d}+})^{n_3}(g_{++})^{n_2}(g_{--})^{n_1}
	\end{equation}	
	where $n_i$'s are the number of each gate. One can verify that 
	\bea
	(g_{--})^{n_1}\,\Psi_R(x_-,x_+)=(\frac{\omega_R}{\pi})^\frac{1}{2}\,e^{n_1\epsilon/2}\exp\Big[-\frac{\omega_R}{2} e^{2n_1\epsilon}x_-^2-\frac{\omega_R}{2}x_+^2\Big],
	\eea
	also one finds
	\bea
	(g_{++})^{n_2}\,(g_{--})^{n_1}\,\Psi_R(x_-,x_+)=(\frac{\omega_R}{\pi})^\frac{1}{2}\,e^{(n_1+n_2)\epsilon/2}\exp\Big[-\frac{\omega_R}{2} e^{2n_1\epsilon}x_-^2-e^{2n_2\epsilon}\frac{\omega_R}{2}x_+^2\Big],
	\eea
	on the other hand, the displacement gate $g_{\mathbf{d}+}$ shifts the $x_+$ coordinates as 
	\bea \label{tar12}
	(g_{\mathbf{d}+})^{n_3}(g_{++})^{n_2}\,(g_{--})^{n_1}\,\Psi_R(x_-,x_+)=(\frac{\omega_R}{\pi})^\frac{1}{2}\,e^{(n_1+n_2)\epsilon/2}e^{(-\frac{\omega_R}{2}\epsilon^2n_3^2\,\mathbf{d}^2\,e^{2n_2\epsilon})}\nonumber\\\hspace*{3cm}\exp\Big[-\frac{\omega_R}{2} e^{2n_1\epsilon}x_-^2-e^{2n_2\epsilon}\frac{\omega_R}{2}x_+^2-e^{2n_2\epsilon}\omega_R\epsilon \,n_3\mathbf{d}\,x_+\Big].
	\eea
	From \eqref{tar12} and \eqref{tar123} one obtains
	\bea
	&&\omega_R e^{2n_1\epsilon}=\omega_-e^{2\alpha},\nonumber\\
	&&\omega_R e^{2n_2\epsilon }=\omega_+\,e^{-2\alpha},\nonumber\\
	&& e^{2n_2\epsilon}\omega_R\epsilon \,n_3=-\sqrt{2} \omega_+e^{-2\alpha}
	\eea
	leading to 
	\bea n_1&=&\frac{1}{\epsilon}\Big(\alpha+\frac{1}{2}\log \frac{\omega_-}{\omega_R}\Big),\nonumber\\
	n_2&=&\frac{1}{\epsilon}\Big(-\alpha+\frac{1}{2}\log \frac{\omega_+}{\omega_R}\Big),\nonumber\\
	n_3&=&-\frac{\sqrt{2}}{\epsilon},
	\eea
	these also satisfy the normalization factor of the target state. Therefore, by definition, the depth of the circuit is given by  
	\bea
	C(\hat U)&=&\sum_{n_i}\,|n_i|\nonumber\\
	&=&\frac{1}{2\epsilon}\bigg(|\log \frac{\omega_-}{\omega_R}+2\alpha|+|\log \frac{\omega_+}{\omega_R}\,-2\alpha|+2\sqrt{2}\bigg),
	\eea
	where $\alpha$ is given by \eqref{al}. One may highlight that the statement of $	C(\hat U)$ does not necessarily imply an optimal circuit and that Nielsen’s approach offers a method to derive a circuit complexity.	Additionally, the construction of the $\hat U$ operator demonstrated the necessity of the displacement gate, despite the absence of the displacement $\mathbf{d}$ in the circuit depth.
	%--------------------------------------------------------------------------
	\section{Eigenvalues of the $	{\cal M}_{+}$ }\label{appC}
	
	The eigenvalues may be found by solving the following determinant equation 
	\bea
	\mid\hspace{-1mm}	{\cal M}_{+}-	{\cal M}^{(i)}_{+} I\hspace{-1mm}\mid=0,
	\eea
	to solve the above equation, one may use the perturbative expansion for the eigenvalues
	\bea\label{EigenDeltapp}
	{\cal M}^{(i)}_{+}=	({\cal M}^{(i)}_{+})_0+	({\cal M}^{(i)}_{+})_1\hspace{.5mm}\tilde{\mathbf{d}}^2+...,\hspace{1cm}\hspace{3mm}i=1,2,3
	\eea
	After simplification, one can see that
	\bea
	({\cal M}^{(i)}_{+})_0= 	{\cal M}^{(i)}_{-}\hspace{.5mm}\left(\alpha\rightarrow -\alpha\right),
	\eea
	where  $	{\cal M}^{(i)}_{-}$ are given by \eqref{EigenDeltamm}
	and
	\bea
	&&	({\cal M}^{(1)}_{+})_1=\frac{4\lambda\lambda_{R}\bigg(\lambda\hspace{.5mm} \cosh 2\alpha-\lambda\cos\omega\,t\,\sinh 2\alpha-\lambda_R\bigg)}{(\lambda^2+\lambda_R^2)\cosh2\alpha +(\lambda_R^2 -\lambda^2)\cos\omega t\,\sinh2\alpha -2\lambda \lambda _R},
	\cr\nonumber\\
	&&	({\cal M}^{(2)}_{+})_1=\frac{4\hspace{.5mm} \lambda^{3}_{R}\hspace{1mm} \Big(\cosh 2\alpha +\cos \omega t \,\sinh 2\alpha \Big)\hspace{.5mm}	{\cal M}^{(2)}_{0,+}- 4\hspace{.5mm} \hspace{.5mm}\lambda_{R}^2\,\lambda }{\Big((\lambda^2+\lambda_R^2)\cosh2\alpha +(\lambda_R^2 -\lambda^2)\cos\omega t\,\sinh2\alpha -2\lambda \lambda _R\Big)}\frac{1}{\Big(1+\hspace{.5mm}	({\cal M}^{(2)}_{+})_0\Big)},
	\cr \nonumber\\
	&&	({\cal M}^{(3)}_{+})_1=\frac{4\hspace{.5mm} \lambda^3_{R}\hspace{1mm} \Big(\cosh 2\alpha +\cos \omega t \,\sinh 2\alpha \Big)\hspace{.5mm}	{\cal M}^{(3)}_{0,+}- 4\hspace{.5mm} \hspace{.5mm}\lambda_R^2\lambda }{\Big((\lambda^2+\lambda_R^2)\cosh2\alpha +(\lambda_R^2 -\lambda^2)\cos\omega t\,\sinh2\alpha -2\lambda \lambda _R\Big)}\frac{1}{\Big(1+\hspace{.5mm}	({\cal M}^{(3)}_{+})_0\Big)}.
	\eea
	
	%\section{Appendix}

\end{document}